\documentclass[aps,prd,10pt,nofootinbib,twocolumn,eqsecnum,showpacs,showkeys,superscriptaddress,preprintnumbers,altaffilletter]{revtex4-2}
\usepackage{graphicx}
\usepackage{hyperref}
\usepackage{subcaption}
\usepackage{amsmath}
\usepackage{multirow}
\usepackage{makecell}
\usepackage{dcolumn}
\usepackage{amssymb}
\usepackage{amsfonts}
\usepackage{amsbsy}

\begin{document}

\title{One Merger, Three Bands: Multiband Detection Rates and Parameter Estimation of Stellar-mass Binary Black Hole Coalescences from Millihertz to Kilohertz}

\date{\today}

\author{Anson Chen}
\email{chena@ucas.ac.cn}  
% \altaffiliation{chena@ucas.ac.cn}
\affiliation{International Center for Theoretical Physics Asia-Pacific, University of Chinese Academy of Sciences, 100190 Beijing, China}

\author{Jun Zhang}
\email{zhangjun@ucas.ac.cn}
\affiliation{International Center for Theoretical Physics Asia-Pacific, University of Chinese Academy of Sciences, 100190 Beijing, China}
\affiliation{Taiji Laboratory for Gravitational Wave Universe, University of Chinese Academy of Sciences, 100049 Beijing, China}

\begin{abstract}
Stellar-mass binary black holes sweep through the millihertz, decihertz, and kilohertz gravitational-wave bands before merger, making them prime targets for multiband observations. We forecast their detection rates and parameter-estimation precision for networks combining the millihertz observatories LISA, Taiji, and TianQin; the decihertz concepts LGWA, AMIGO, and AMIGO-5; and the ground-based LVK and Cosmic Explorer detectors. Using the binary-black-hole mass and redshift distributions inferred from GWTC-4, we first estimate the single-band yields and then account explicitly for detector lifetimes and relative mission start times when calculating multiband rates. For a signal-to-noise-ratio threshold of 8, the LISA--Taiji--TianQin network is expected to detect $53.6^{+19.7}_{-14.1}$ systems in the millihertz band, while the largest three-band yield is $23.5^{+8.7}_{-6.2}$ events for LISA--Taiji--TianQin combined with LGWA. The yield is maximized when the decihertz observation begins approximately $0$--$3\,\mathrm{yr}$ after the millihertz mission. Ground-informed subthreshold searches with a threshold of 5 increase the multiband yields by a factor of approximately $4$--$5$, reaching $125.1^{+46.1}_{-32.9}$ events for the same network. Fisher-matrix forecasts for a GW150914-like source show that decihertz observations provide the dominant improvement in most multiband parameter constraints, whereas millihertz data primarily sharpen the detector-frame chirp mass and sky localization. Combining the three bands improves the chirp-mass precision by orders of magnitude relative to ground-based observations alone and can reduce the localization area and coalescence-time uncertainty substantially. These results demonstrate that coordinated mission scheduling and targeted subthreshold searches are essential for realizing the scientific potential of stellar-mass multiband gravitational-wave astronomy.
\end{abstract}

\maketitle

\section{Introduction}

Since the first direct detection of gravitational waves by the Laser Interferometer Gravitational-Wave Observatory (LIGO) in 2015 \cite{LIGOScientific:2016aoc}, we have entered the era of gravitational-wave (GW) astronomy. A decade later, the ground-based LIGO–Virgo–KAGRA (LVK) detector network completed its fourth observing run (O4), and the latest Gravitational-Wave Transient Catalog 5.0 (GWTC-5.0) from the LVK contains
390 GW events originating from compact binary coalescences (CBCs) \cite{LIGOScientific:2026sit,LIGOScientific:2026wfs}. These detections provide unique laboratories for investigating compact-object populations \cite{LIGOScientific:2026vua,LIGOScientific:2026ctl}, testing theories of gravity \cite{LIGOScientific:2026qni,LIGOScientific:2026fcf,LIGOScientific:2026wpt,LIGOScientific:2026oim}, probing cosmology \cite{LIGOScientific:2025jau,LIGOScientific:2026uyd,Chen:2026owi}, and addressing a broad range of fundamental questions in astrophysics and physics. 

The promise of GW astronomy will continue to unfold over the coming decades if new generations of GW detectors come online in the near future. Next-generation ground-based observatories, such as the Einstein Telescope (ET) \cite{Punturo:2010zz,Branchesi:2023mws} and the Cosmic Explorer (CE)
\cite{Evans:2021gyd,Srivastava:2022slt,Evans:2023euw}, will substantially improve sensitivity in the frequency range of $10$--$10^3\,\mathrm{Hz}$. Meanwhile, many GW observatories targeting other frequency bands have been proposed. Space-based detectors such as LISA (Laser Interferometer Space Antenna) \cite{amaroseoane2017,Robson:2018ifk,Baker:2019nia,Babak:2021mhe,Bayle:2022hvs}, Taiji \cite{Ruan:2018tsw,Ruan:2020smc,Gong:2021gvw,Liu:2023qap}, and TianQin
\cite{TianQin:2015yph,Luo:2020bls,TianQin:2020hid,Gong:2021gvw} are designed to detect GWs in the millihertz band. Several space-based missions targeting the decihertz band have also been proposed, including DECIGO (Decihertz Interferometer Gravitational-Wave Observatory) \cite{Kawamura:2006up,Kawamura:2020pcg}, BBO (Big Bang Observer) \cite{Crowder:2005nr,2006CQGra..23.4887H,Cutler:2005qq}, TianGO \cite{Kuns:2019upi}, and AMIGO (Astrodynamical Middle-frequency Interferometric Gravitational-Wave Observatory) \cite{Ni:2019nau,Ni:2021eqz}. 
Furthermore, several lunar-based GW detectors targeting the decihertz band have been proposed in recent years, including LGWA (Lunar Gravitational-wave Antenna) \cite{LGWA:2020mma,Ajith:2024mie,Iacovelli:2025kwn}, LILA (Laser interferometer lunar antenna) \cite{Jani:2025uaz}, CIGO (Crater Interferometry Gravitational-wave Observatory) \cite{Zhang:2025eeh}, and LANGO (Lunar accelerometer network gravitational observatory) \cite{2026CQGra..43i5023P}. In addition to conventional Michelson interferometry, atom-interferometric GW detectors aiming at the decihertz band have also been proposed. Notable examples include MAGIS (Mid-band Atomic Gravitational
Wave Interferometric Sensor) \cite{Graham:2012sy,Graham:2016plp,Graham:2017pmn,Coleman:2018ozp}, AION (Atom Interferometer Observatory and Network) \cite{Badurina:2019hst}, ELGAR (European Laboratory for Gravitation and Atom-interferometric Research) \cite{Canuel:2019abg}, ZAIGA (Zhaoshan Long-baseline Atom Interferometer Gravitation Antenna) \cite{Zhan:2019quq}, and so on. Together, these diverse detector concepts promise a bright future for multiband GW astronomy.
\begin{figure*}[t]
    \centering
    \includegraphics[width=0.8\linewidth]{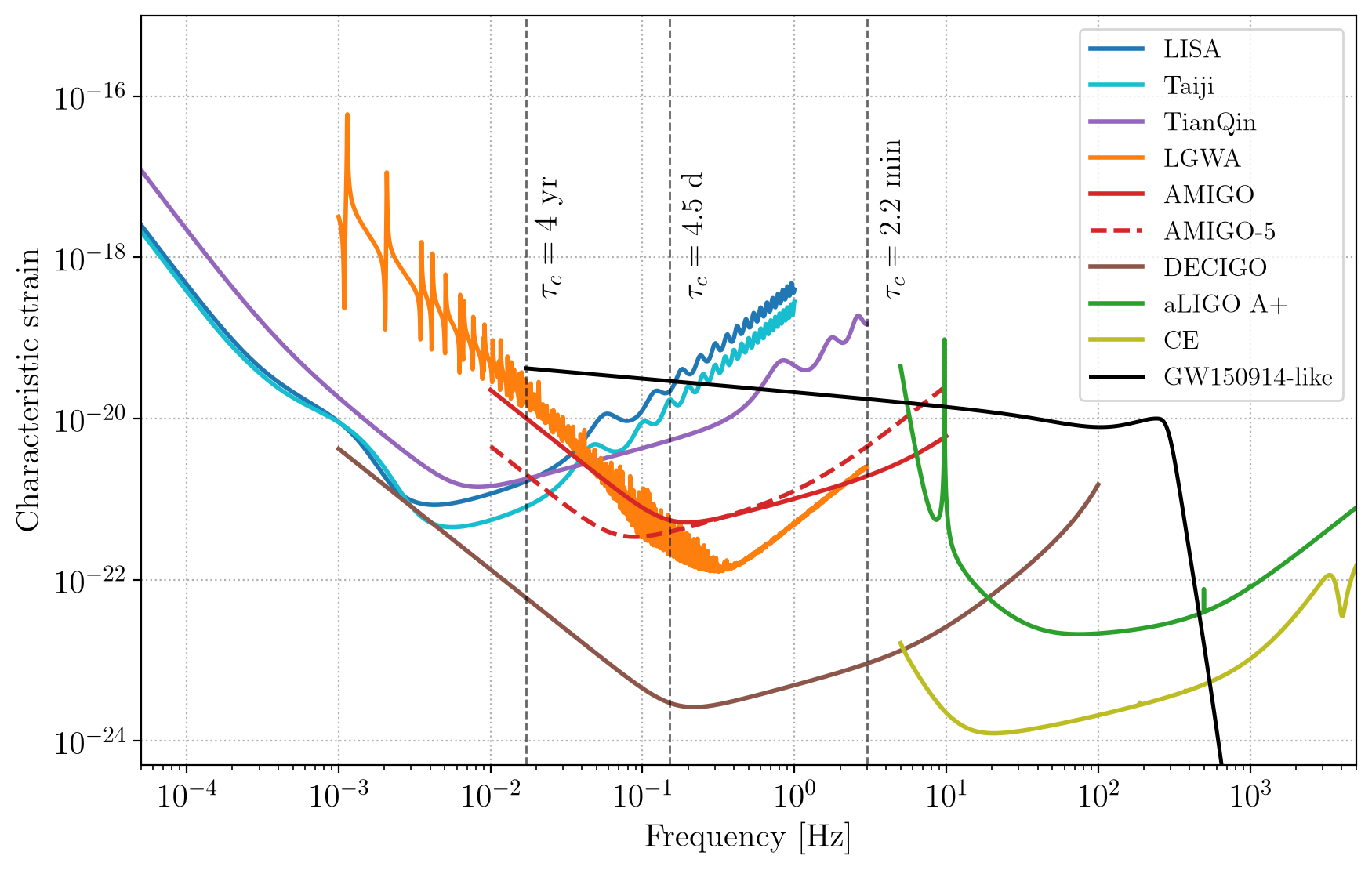}
    \caption{Characteristic strain of a GW150914-like event compared with the sensitivity curves of detectors operating in different frequency bands.}
    \label{fig:GW150914_PSD}
\end{figure*}

In principle, all of the GW detectors discussed above can observe signals from stellar-mass binary black hole (SBBH) coalescences, whose inspiral–merger–ringdown signals evolve from the millihertz to the kilohertz frequency band. If multiple detectors operate during overlapping periods, the same event could be tracked from its early inspiral in the millihertz band several years before coalescence, through the decihertz band, and finally to its merger in the kilohertz band, as shown in Fig.~\ref{fig:GW150914_PSD}. Such multiband observations would provide powerful probes of compact-object formation channels \cite{Gerosa:2019dbe}, improve dark-siren cosmography through enhanced source localization \cite{Muttoni:2021veo,Seymour:2022teq,Dong:2024bvw,Dong:2026uxr,Song:2026kii,Song:2026tdy}, and enable stringent tests of gravity \cite{Carson:2019kkh,Datta:2020vcj,Jia:2026vqo}, owing to their long observational baselines and broad frequency coverage.

In particular, comparing the coalescence times inferred independently from signals in different frequency bands offers a sensitive test of fundamental physics. For example, in certain effective field theories (EFTs) of modified gravity, the propagation speed of GWs may transition from the speed of light to a different value near the ultraviolet cutoff scale \cite{deRham:2018red,LISACosmologyWorkingGroup:2022wjo}. Such frequency-dependent propagation effects could be constrained through multiband GW observations \cite{Baker:2022eiz,Harry:2022zey,Chen:2024ery,Praveen:2025nqj}. Moreover, superradiant axion clouds surrounding black holes may modify the GW waveform during the early inspiral \cite{Baumann:2019eav,Takahashi:2021yhy,Takahashi:2021eso,Takahashi:2024fyq}. Gravitational perturbations from the companion can subsequently induce transitions between the cloud’s energy levels, causing the late-inspiral waveform to approach the prediction of general relativity. Multiband observations are therefore essential for probing the characteristic frequency-dependent signatures of axion-cloud superradiance \cite{Peng2026}.

Previous studies have forecast the multiband detection rates of SBBHs jointly observed by LISA-like and ground-based detectors \cite{Sesana:2017vsj,Wong:2018uwb,Gerosa:2019dbe}, or by LGWA and ground-based detectors \cite{Iacovelli:2025kwn}. The prospects for detecting SBBHs across all three frequency bands have also been explored using a simplified approach in which the accumulated signal-to-noise ratio (SNR) is computed from a fixed initial frequency, without accounting for the distribution of times to coalescence for different events \cite{Zhao:2023ilw}. In addition, despite the substantial uncertainty in their underlying population, the multiband detection of intermediate-mass binary black hole coalescences has also been investigated \cite{Dong:2025ikq}.
In this work, we forecast the number of SBBH coalescences detectable by multiband networks comprising one or more millihertz space-based observatories—LISA, Taiji, and TianQin; decihertz observatories represented by the lunar-based LGWA and the space-based AMIGO; and ground-based detectors, including the LVK network at design sensitivity and the next-generation network represented by two Cosmic Explorers located at the LIGO sites. Compared with previous studies, we adopt the updated stellar-mass black hole population model inferred from GWTC-4 \cite{LIGOScientific:2026vua,LIGOScientific:2025jau}. In particular, we investigate for the first time how the detection number depends on the time delay between the observing periods of millihertz and decihertz detectors. We find that the expected number of detections is sensitive to this time gap.

The remainder of this paper is organized as follows. In Sec.\ref{sec:single_band}, we first forecast the numbers of events detectable by different instruments in the millihertz and decihertz frequency bands. In Sec.\ref{sec:multi_band}, we then investigate multiband detections for different time
delays between the observation periods of millihertz and decihertz detectors. In addition, in Sec.\ref{sec:pe}, we assess the parameter-estimation capabilities of multiband observations using the Fisher-matrix formalism. Finally, we summarize our conclusions in Sec.\ref{sec:conclusion}.

\section{Single-band detection of SBBHs}
\label{sec:single_band}

In this section, we forecast the number of SBBHs detectable in the frequency bands targeted by different GW observatories. 

\subsection{Detectability}
\label{sec:detectability}

During the inspiral, the GW frequency increases as the binary approaches coalescence. Because multiband observations probe the early inspiral, the frequency evolution can be adequately described at leading order in terms of the time to coalescence, $\tau_c$, as \cite{Gerosa:2019dbe}
\begin{equation}
  f(\tau_c) = \frac{5^{3/8}}{8\pi}
  \left[{\cal M}_c(1+z)\right]^{-5/8}\,\tau_c^{-3/8},
\end{equation}
where $z$ is the source redshift, and ${\cal M}_c$ is the source-frame chirp mass defined by the component masses $m_1$ and $m_2$ as ${\cal M}_c={(m_1m_2)^{3/5}}/{(m_1+m_2)^{1/5}}$. Geometric units $G=c=1$ are used throughout the paper. Consequently, signals observed at lower frequencies generally remain in band for longer periods before evolving to higher frequencies. Taking a GW150914-like event as an example, its GW frequency is approximately $0.016\,{\rm Hz}$ at $\tau_c=4\,{\rm yr}$. This frequency lies near the high-frequency end of the sensitivity band of LISA-like detectors and the low-frequency end of that of decihertz detectors, as shown in Fig.~\ref{fig:GW150914_PSD}. For LISA-like detectors, most of the SNR from such a source accumulates during the first few years of observation, and the signal evolves beyond their sensitive frequency band several days before coalescence. By contrast, in decihertz detectors, most of the SNR accumulates during the final few days to hours before merger. The late inspiral and merger then occur over only a few seconds in the frequency band of ground-based detectors. 

Notably, space-based and lunar-based detectors operate for a finite observation duration, denoted by $T_{\rm obs}$. If $\tau_c\leq T_{\rm obs}$, the binary coalesces during the observing period, and its signal evolves from the initial frequency $f(\tau_c)$ to the upper limit of the detector’s frequency band, $f_{\rm high}$. If $\tau_c>T_{\rm obs}$, only a portion of the inspiral is observed, with the signal evolving from $f(\tau_c)$ to $f(\tau_c-T_{\rm obs})$. The SNR $\rho$ accumulated during the observing period is therefore computed as \cite{Gerosa:2019dbe}
\begin{equation}
  \rho^2(\tau_c) = 4\int_{f(\tau_c)}^{f_{\rm end}}
  \frac{|\tilde{h}(f)|^2}{S_n(f)}\,{\rm d}f,
  \label{eq:SNR_tc}
\end{equation}
where $\tilde{h}(f)$ is the frequency-domain GW waveform, $S_n(f)$ is the detector noise power spectral density (PSD), and the upper frequency bound $f_{\rm end}$ is given by
\[
f_{\rm end}=
\begin{cases}
f_{\rm high},&\tau_c\leq T_{\rm obs},\\
\min[f(\tau_c-T_{\rm obs}),f_{\rm high}],&\tau_c>T_{\rm obs}.
\end{cases}
\]
Because the SNRs in the frequency bands of space- and lunar-based detectors accumulate primarily during the early inspiral, higher-order frequency-dependent corrections that become important near merger can be safely neglected. We therefore use the leading-order stationary-phase inspiral amplitude, given by \cite{Ajith:2007kx},
\begin{equation}
    |\tilde{h}(f)| = \sqrt{\frac{5}{24}}{\pi^{-2/3}} \frac{\left[{\cal M}_c(1+z)\right]^{5/6}}{D_L} f^{-7/6},
\end{equation}
where $D_L$ denotes the luminosity distance of the source. Throughout the paper, we adopt the $\Lambda$CDM cosmology model with $H_0=67.66\,{\rm km}\,{\rm s}^{-1}\,{\rm Mpc}^{-1}, \Omega_{m0} = 0.30966$ from Planck 18 \cite{Planck:2018vyg} for computing $D_L(z)$. 

For the space-based detectors, we use the sky-averaged sensitivity curves and response function presented in Ref.~\cite{Robson:2018ifk}, adopting the corresponding detector-specific parameters, as shown in Fig.~\ref{fig:GW150914_PSD}. For LGWA, we use the sensitivity curve implemented in the \texttt{GWFish} package \cite{Dupletsa:2022scg,Iacovelli:2025kwn}, which models the detector as two orthogonal lunar-array components. The antenna response of each component is described by the polarization functions $F_+$ and $F_\times$ of a right-angle Michelson interferometer. The sky-averaged LGWA SNR, $\langle\rho^2\rangle$, therefore incorporates the antenna-pattern averages $\langle F_+^2\rangle=\langle F_\times^2\rangle=1/5$ \cite{Maggiore:2007ulw}. For all detectors, we additionally include a factor of $4/5$ in $\rho^2$ to account for averaging over the binary inclination angle.

\begin{figure}
    \centering
    \includegraphics[width=\linewidth]{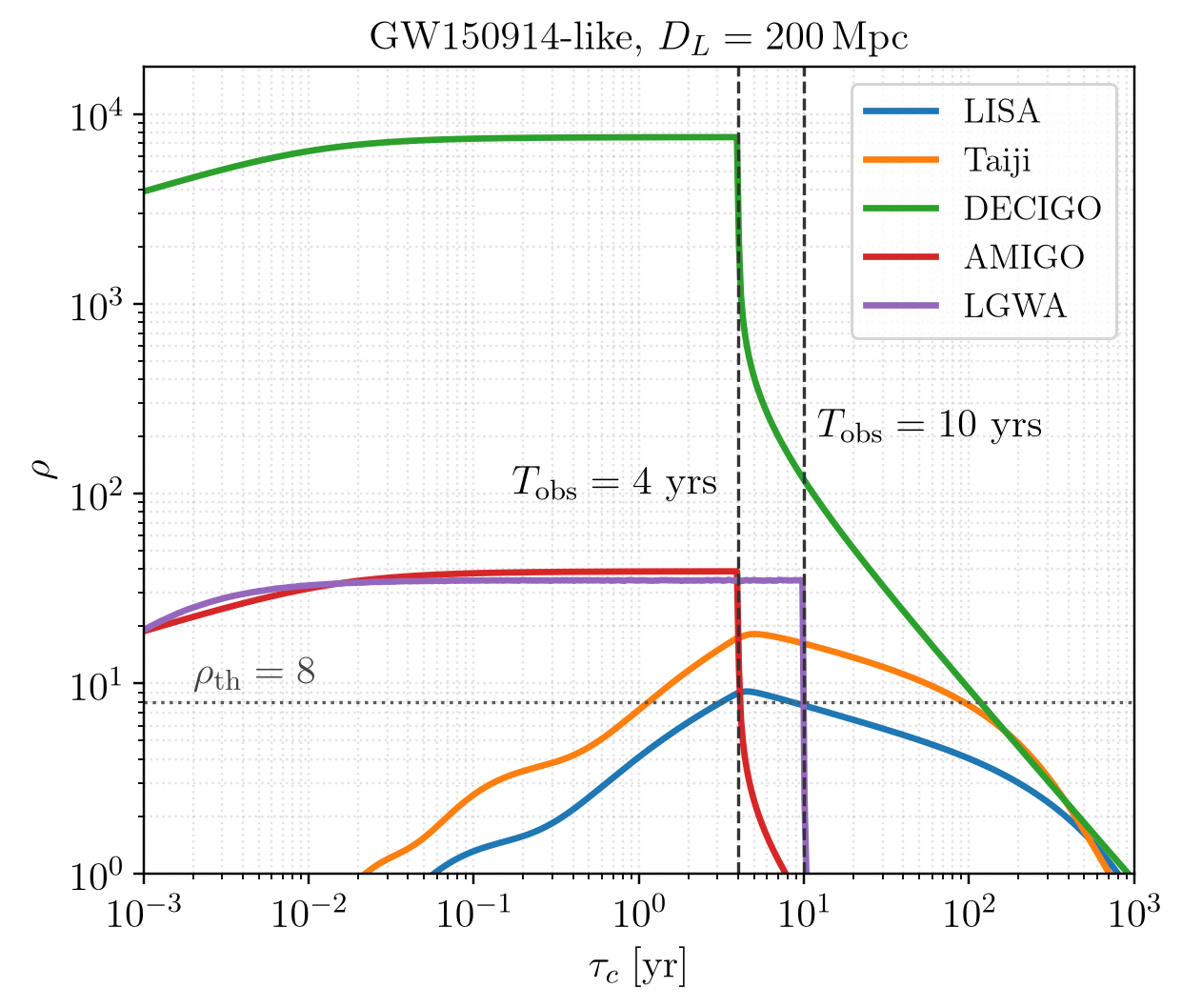}
    \caption{Sky-averaged SNR of a GW150914-like event located at $D_L=200\,{\rm Mpc}$ as a function of $\tau_c$ for different space-based detectors and LGWA.}
    \label{fig:SNR_tc}
\end{figure}

Fig.~\ref{fig:SNR_tc} shows the sky-averaged SNR of a GW150914-like event located at a luminosity distance of $D_L=200\,{\rm Mpc}$, computed using Eq.~\eqref{eq:SNR_tc} as a function of $\tau_c$ for different space-based detectors and LGWA. The SNR curves indicate that millihertz detectors cannot detect sources with very short times to coalescence because their signals have already evolved beyond the most sensitive part of the detector band. However, the maximum detectable $\tau_c$ can extend to tens or even hundreds of years. The SNR generally peaks near $\tau_c=T_{\rm obs}$, where we assume $T_{\rm obs}=4\,{\rm yr}$ for the space-based detectors.
On the other hand, the SNRs of decihertz detectors remain high at small $\tau_c$ but decrease rapidly once $\tau_c$ exceeds $T_{\rm obs}$, because most of the SNR accumulates during the final few days of the inspiral. For AMIGO and LGWA, the SNR falls below the detection threshold $\rho_{\rm th}=8$ shortly after $\tau_c$ exceeds $T_{\rm obs}$ (Adopting $T_{\rm obs}=10\,{\rm yr}$ for LGWA). Owing to its substantially better design sensitivity, however, DECIGO maintains an SNR well above the threshold and can detect sources hundreds of years before coalescence.

\begin{figure}
    \centering
    \includegraphics[width=\linewidth]{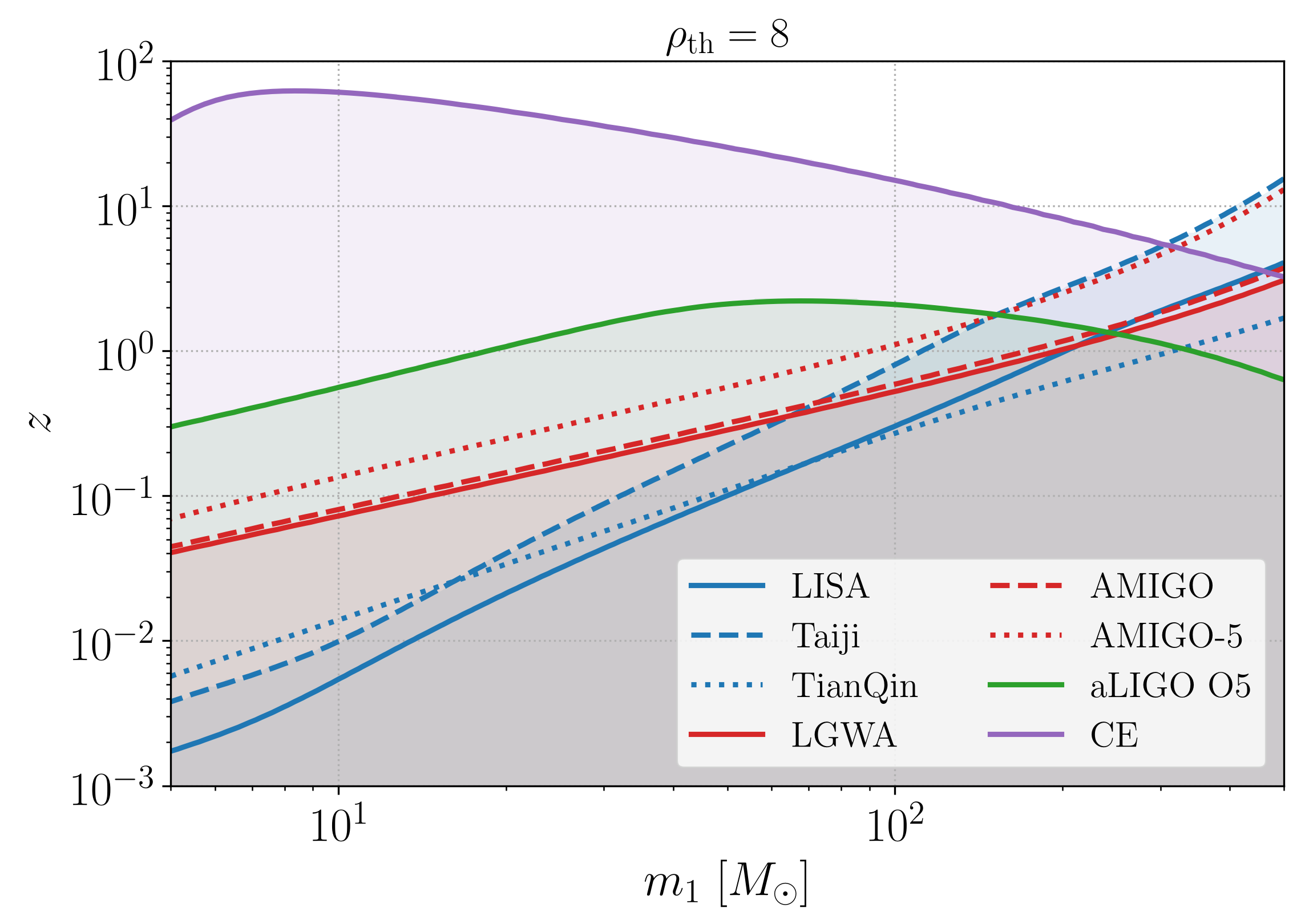}
    \caption{Regions in the $(m_1,z)$ parameter space where the sky-averaged SNR of equal-mass, nonspinning binaries exceeds the detection threshold $\rho_{\rm th}=8$, indicated by the shaded areas.}
    \label{fig:SNR_contour}
\end{figure}

In addition, Fig.~\ref{fig:SNR_contour} shows the regions in the component-mass–redshift plane where equal-mass binaries have sky-averaged SNRs above the detection threshold $\rho_{\rm th}=8$. We compare the detection reach of millihertz, decihertz, and ground-based detectors. For the space-based detectors and LGWA, we set $\tau_c=T_{\rm obs}$, which approximately maximizes the SNR. For the ground-based detectors, we calculate the full inspiral–merger–ringdown SNR with the PhenomA waveform \cite{Ajith:2007kx}, integrating from the lower frequency cutoff of each detector and adopting the design sensitivities of Advanced LIGO and the 40-km Cosmic Explorer. 
As shown in Fig.~\ref{fig:SNR_contour}, millihertz detectors have the smallest redshift reach for stellar-mass systems with $m_1<100\,M_\odot$, typically of order $z\sim10^{-1}$. Among these detectors, Taiji generally achieves the greatest reach because of its better sensitivity near the signals’ initial frequencies, although TianQin performs slightly better for low-mass systems owing to its superior sensitivity at higher frequencies. AMIGO and LGWA typically reach greater distances than the millihertz detectors; however, for $m_1\gtrsim80\,M_\odot$, Taiji outperforms both. AMIGO-5, an alternative design with an arm length of $50{,}000\,{\rm km}$ rather than $10{,}000\,{\rm km}$ adopted for the original AMIGO \cite{Ni:2021eqz}, provides the greatest reach among the considered decihertz detectors for $m_1<100\,M_\odot$. Ground-based detectors nevertheless have the greatest overall reach: Advanced LIGO can detect most stellar-mass systems at $z\sim1$, while Cosmic Explorer can extend well beyond $z\sim10$. These results indicate that SBBHs observed by millihertz or decihertz detectors will generally lie within the sensitivity range of ground-based detectors, provided that the latter are operating when the binaries merge. Moreover, given the substantially higher SNRs achieved by ground-based detectors, its changes due to corrections from more sophisticated waveform models are expected to have little effect on the predicted number of multiband detections.

\subsection{Population model}
\label{sec:population}

To obtain reliable forecasts for the number of detectable SBBHs, we employ the most recent population model derived from the LVK GWTC-4 catalog \cite{LIGOScientific:2026vua,LIGOScientific:2025jau}. The primary-mass distribution of stellar-mass black holes can be described by the Power-law + Multi-peak model:
\begin{align}
    p(m_1) &= (1-\lambda_{\rm g}){\cal B}(m_1|M_{\rm min},M_{\rm max},\alpha) \nonumber\\
    &\quad +\lambda_{\rm g}\lambda_{\rm g,low}
    {\cal G}(m_1|\mu_{\rm g,low},\sigma_{\rm g,low}) \nonumber\\
    &\quad +\lambda_{\rm g}(1-\lambda_{\rm g,low})
    {\cal G}(m_1|\mu_{\rm g,high},\sigma_{\rm g,high}),
\end{align}
where ${\cal B}$ represents the broken-power-law component and ${\cal G}$ denotes a Gaussian distribution. The model contains ten parameters: the lower and upper mass bounds, $M_{\rm min}$ and $M_{\rm max}$; the power-law index $\alpha$; the combined contribution of the Gaussian components, $\lambda_{\rm g}$; the relative weight of the lower-mass Gaussian, $\lambda_{\rm g,low}$; the Gaussian means, $\mu_{\rm g,low}$ and
$\mu_{\rm g,high}$; the corresponding standard deviations, $\sigma_{\rm g,low}$ and $\sigma_{\rm g,high}$; and the low-mass smoothing scale, $\delta_m$. The secondary mass is specified through a power-law mass-ratio distribution with index $\beta$, which, together with the primary-mass distribution, defines the joint component-mass distribution $p(m_1,m_2)$. For our forecasts, we adopt the maximum-likelihood population-parameter values inferred from the GWTC-4 data, as summarized in
Table~\ref{tab:GWTC4_param}. 

The redshift distribution of SBBHs is determined by the intrinsic merger-rate density and the comoving-volume element:
\begin{equation}
    p(z) \propto {\cal R}(z)\frac{{\rm d}V_c}{{\rm d}z}\frac{1}{1+z},
\end{equation}
where the factor $1/(1+z)$ accounts for cosmological time dilation between the source and observer frames. Assuming that all SBBHs arise from stellar progenitors, their merger-rate evolution is expected to approximately follow the cosmic star-formation history. We model this evolution using the Madau--Dickinson parameterization \cite{Madau:2014bja}:
\begin{equation}
    {\cal R}(z) = R_0 \left[1+(1+z_p)^{-\gamma-k}\right] \frac{(1+z)^\gamma}{1+\left(\dfrac{1+z}{1+z_p}\right)^{\gamma+k}},
\end{equation}
where $R_0$ denotes the local merger-rate density at $z=0$. The GWTC-4 analysis constrains the BBH rate to be $R_0=19^{+7}_{-5}\,{\rm Gpc}^{-3}\,{\rm yr}^{-1}$ \cite{LIGOScientific:2026vua}. The parameters $\gamma$ and $k$ describe the power-law evolution below and above the turnover
redshift $z_p$ respectively, and we adopt their maximum-likelihood values inferred from GWTC-4 as listed in Table~\ref{tab:GWTC4_param}.
\begin{table}
    \centering
    \caption{Maximum-likelihood parameters for the Powerlaw + Multipeak mass distribution model and the Madau-Dickinson merger rate redshift evolution model inferred from GWTC-4.}
    \label{tab:GWTC4_param}
    \begin{tabular}{llll}
        \hline
        Parameter & ~~~~~~Value & ~~~~~~Parameter & ~~~~~~Value \\
        \hline
        $\alpha$ & $~~~~~~2.9$ & $~~~~~~\mu_{\rm g,low}[M_\odot]$ & $~~~~~~9.7$ \\
        $\beta$ & $~~~~~~1.0$ & $~~~~~~\mu_{\rm g,high}[M_\odot]$ & $~~~~~~30.7$ \\
        $M_{\rm min}[M_\odot]$ & $~~~~~~4.6$ & $~~~~~~\sigma_{\rm g,low}[M_\odot]$ & $~~~~~~0.7$ \\
        $M_{\rm max}[M_\odot]$ & $~~~~~~86.3$ & $~~~~~~\sigma_{\rm g,high}[M_\odot]$ & $~~~~~~6.3$ \\ 
        $\lambda_{\rm g}$ & $~~~~~~0.4$ & $~~~~~~\gamma$ & $~~~~~~3.3$ \\
        $\lambda_{\rm g,low}$ & $~~~~~~0.8$ & $~~~~~~\kappa$ & $~~~~~~2.9$ \\
        $\delta_m[M_\odot]$ & $~~~~~~4.8$ & $~~~~~~z_p$ & $~~~~~~2.5$ \\
        \hline
    \end{tabular}
\end{table}

\subsection{Detection number and parameter distribution}

To forecast the number of detectable long-lived SBBHs, we must account for all systems whose SNR exceeds the detection threshold at any time during the observation period. As discussed in Sec.~\ref{sec:detectability}, each source is detectable within a time window bounded by the threshold-crossing points satisfying $\rho(\tau_c)=\rho_{\rm th}$. As illustrated in Fig.~\ref{fig:SNR_tc}, LISA-like detectors generally exhibit two such crossings, yielding a detection window of $|\tau_{\rm th,1}(m_1,m_2,z)-\tau_{\rm th,2}(m_1,m_2,z)|$, where $\tau_{\rm th,1}$ and $\tau_{\rm th,2}$ denote the times to coalescence at the two crossings and depend on the source masses and redshift. For decihertz detectors, the lower threshold-crossing time can be very short, potentially occurring only minutes before merger. Assuming that SBBH coalescence times are uniformly distributed, the expected number of detections is given by \cite{Gerosa:2019dbe}
\begin{align}
    N = & \iiint {\rm d}m_1{\rm d}m_2{\rm d}z\,p(m_1,m_2){\cal R}(z)\frac{{\rm d}V_c(z)}{{\rm d}z}\frac{1}{1+z} \nonumber\\
    & \times \bigg|\tau_{\rm th,1}(m_1,m_2,z)-\tau_{\rm th,2}(m_1,m_2,z)\bigg|.
    \label{eq:N_space}
\end{align}
Using the mass and redshift distributions introduced in Sec.\ref{sec:population}, we evaluate the expected number of SBBH detections from Eq. \eqref{eq:N_space} via Monte Carlo integration for the space-based detectors and LGWA. We report both the central estimate and the uncertainty arising from the $90\%$ credible interval of the local merger-rate density from GWTC-4, $R_0=19^{+7}_{-5}\,{\rm Gpc}^{-3}\,{\rm yr}^{-1}$.

For millihertz observations, we consider LISA, Taiji, and TianQin operating individually, as well as the LISA--Taiji, LISA--TianQin, and three-detector LISA--Taiji--TianQin networks. We assume an observation time of $T_{\rm obs}=4\,{\rm yr}$ for each detector. The resulting detection numbers are summarized in Table~\ref{tab:space_number}, showing that LISA is expected to detect a handful of SBBHs, whereas Taiji could detect approximately $30$ systems owing to its greater detection range, as illustrated in Fig.~\ref{fig:SNR_contour}. By contrast, TianQin is expected to detect only about one system. Meanwhile, combining LISA with either Taiji or TianQin substantially increases the expected number of detections relative to the individual detectors. If fortunately all three detectors could operate simultaneously, their joint network will yield the largest detection count, with approximately $54$ SBBHs expected in the millihertz band.
\renewcommand{\arraystretch}{2}
\begin{table}[t]
    \centering
    \begin{tabular}{|c|c|}
    \hline
        Detector & Detection Number \\
        \hline
        LISA & $4.6^{+1.7}_{-1.2}$ \\
        \hline
        Taiji & $29.6^{+10.9}_{-7.8}$ \\
        \hline
        TianQin & $1.1^{+0.4}_{-0.3}$ \\
        \hline
        LISA--Taiji & $47.5^{+17.5}_{-12.5}$ \\
        \hline
        LISA--TianQin & $5.6^{+2.1}_{-1.5}$ \\
        \hline
        LISA--Taiji--TianQin & $53.6^{+19.7}_{-14.1}$ \\
        \hline
    \end{tabular}
    \caption{Forecasted detection numbers and uncertainties for individual space-based detectors and detector networks comprising LISA, Taiji, and TianQin. }
    \label{tab:space_number}
\end{table}

\begin{figure}
    \centering
    \includegraphics[width=\linewidth]{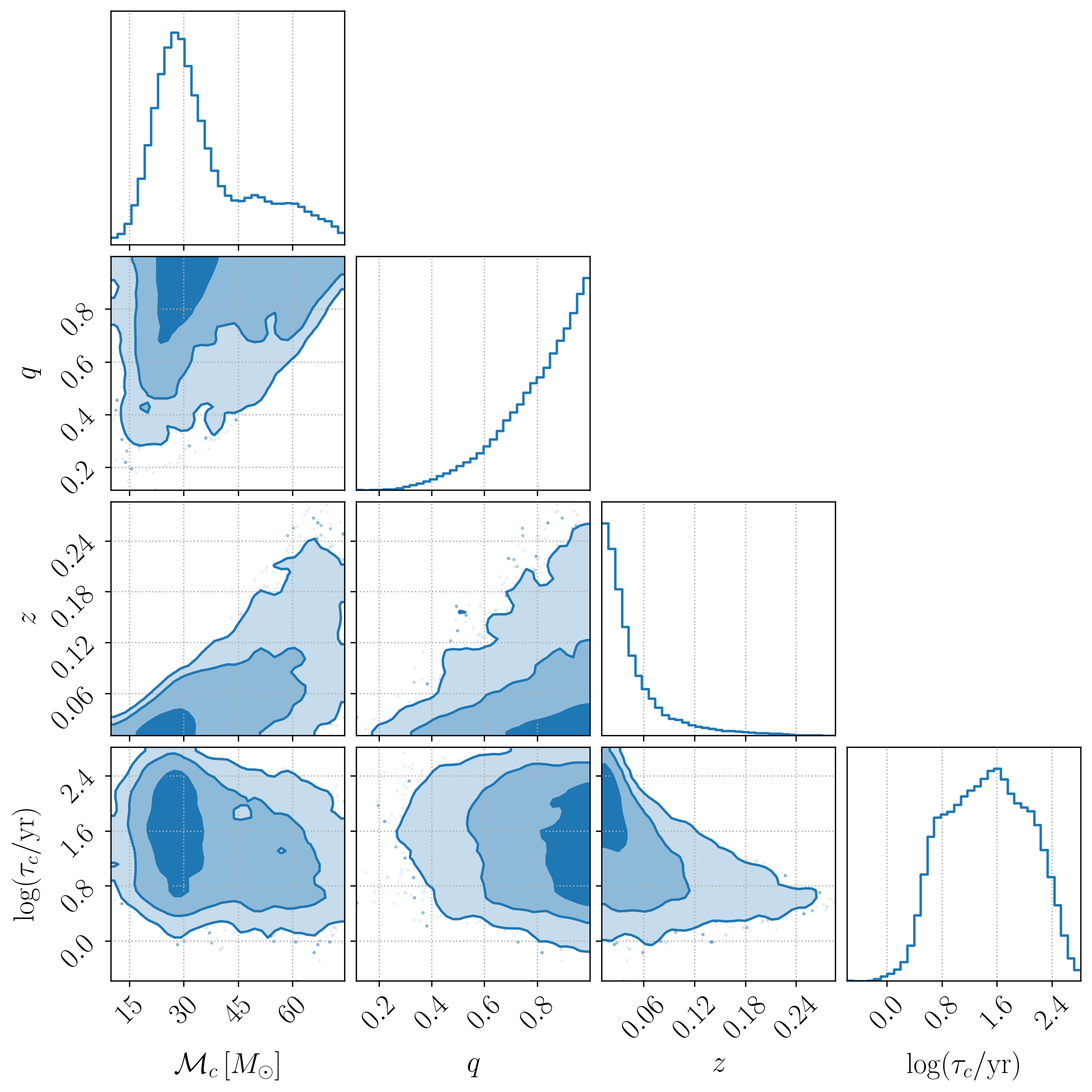}
    \caption{Probability density distribution of the chirp mass, mass ratio, redshift, and logarithm of the time to coalescence for SBBHs detected by LISA with $T_{\rm obs}=4\,{\rm yr}$.}
    \label{fig:lisa_param_dist}
\end{figure}

\begin{figure*}
    \centering
    \includegraphics[width=\linewidth]{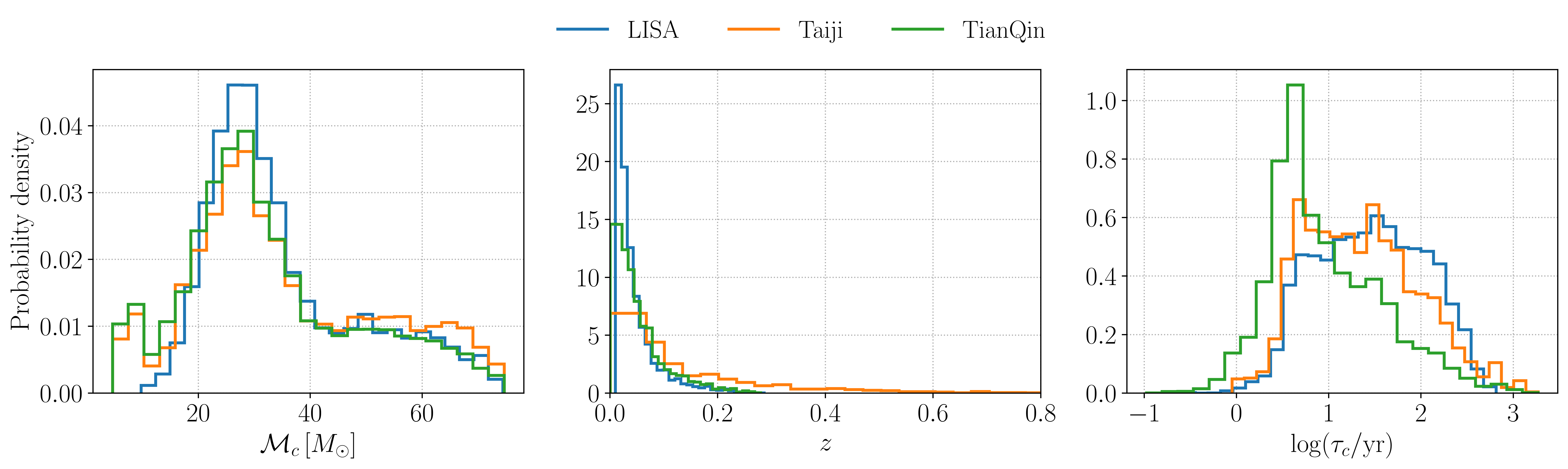}
    \caption{Probability density distribution of the chirp mass, redshift, and logarithm of the time to coalescence for SBBHs detected by the space-based GW observatories LISA, Taiji, and TianQin during a four-year mission.}
    \label{fig:space_param_dist}
\end{figure*}

Using Monte Carlo sampling, we also investigate the parameter distributions of SBBHs detectable by LISA, Taiji, and TianQin. The simulated parameter distributions of the detected SBBHs by LISA are shown in Fig.~\ref{fig:lisa_param_dist}.
% The full corner plot for LISA can be found in our previous work \cite{Peng}. 
In addition, we also compare the one-dimensional probability density distributions of the chirp mass, redshift, and logarithm of the time to coalescence, $\tau_c$, as shown in Fig.~\ref{fig:space_param_dist}.
For all three detectors, the chirp-mass distribution peaks near ${\cal M}_c=30\,M_\odot$, corresponding to the higher-mass Gaussian peak in the adopted mass model. Moreover, the mass-ratio distributions peak at $q=1$ similarly for all detectors as shown in Fig.~\ref{fig:lisa_param_dist}; consequently, the peak of the chirp-mass distribution lies close to that of the primary-mass distribution, consistent with previous forecasts for LISA \cite{Buscicchio:2024asl}. However, Taiji and TianQin detect a larger fraction of low-mass systems with ${\cal M}_c\sim10\,M_\odot$, owing to their slightly greater detection reach for such systems, as illustrated in Fig.~\ref{fig:SNR_contour}.

Furthermore, Fig.~\ref{fig:space_param_dist} shows that most detected events are located at relatively low redshifts, and the number of detections decreases rapidly with increasing redshift. The detected sources for LISA and TianQin are primarily confined to $z<0.3$, whereas Taiji can reach $z\simeq0.8$. This greater redshift reach is the main reason for Taiji’s larger detection yield.
Finally, all three detectors can observe some events as early as $\sim10^3\,{\rm yr}$ before coalescence. The $\tau_c$ distributions for LISA and Taiji are similar, with most events lying between a few years and approximately $100\,{\rm yr}$. By contrast, the distribution for TianQin peaks below $\tau_c=10\,{\rm yr}$, close to the assumed observation duration of $T_{\rm obs}=4\,{\rm yr}$. This behavior likely results from TianQin’s lower sensitivity to these sources, which favors detections near $\tau_c\simeq T_{\rm obs}$, where the SNR is maximized. The higher SNRs achieved by LISA and Taiji broaden the detection window toward larger $\tau_c$, as shown in Fig.~\ref{fig:SNR_tc}, allowing them to detect more events with $\tau_c\gg T_{\rm obs}$.

Turning to decihertz detectors, we forecast the number of SBBHs detectable by LGWA, AMIGO/AMIGO-5, and DECIGO using the same Monte Carlo integration based on Eq.~\eqref{eq:N_space}. We assume an observation duration of $T_{\rm obs}=10\,{\rm yr}$ for LGWA and $T_{\rm obs}=4\,{\rm yr}$ for AMIGO/AMIGO-5, and DECIGO. The resulting detection numbers are presented in Table~\ref{tab:deci_number}.
Although LGWA and AMIGO have comparable detection reaches, as shown in Fig.~\ref{fig:SNR_contour}, LGWA is expected to detect approximately twice as many SBBHs because of its longer observation duration. On the other hand, increasing the arm length of AMIGO-5 to five times that of the original AMIGO design improves its sensitivity and increases the expected detection yield by a factor of approximately five. Nevertheless, owing to its substantially greater design sensitivity, DECIGO has a detection reach exceeding even those of third-generation ground-based detectors and can observe SBBHs beyond $z\sim100$. Its expected detection yield is therefore of order $10^6$, substantially larger than those of the other decihertz detectors and comparable to the yields expected for ET and CE.
\renewcommand{\arraystretch}{2}
\begin{table}[t]
    \centering
    \begin{tabular}{|c|c|}
    \hline
        Detector & Detection Number \\
        \hline
        LGWA & $82.0^{+30.2}_{-21.6}$ \\
        \hline
        AMIGO & $46.2^{+17.0}_{-12.1}$ \\
        \hline
        ~AMIGO-5~ & $256.8^{+94.6}_{-67.6}$ \\
        \hline
        DECIGO & $\sim 10^6$ \\
        \hline
    \end{tabular}
    \caption{Forecasted detection numbers and uncertainties for different decihertz detectors.}
    \label{tab:deci_number}
\end{table}
\begin{figure}
    \centering
    \includegraphics[width=\linewidth]{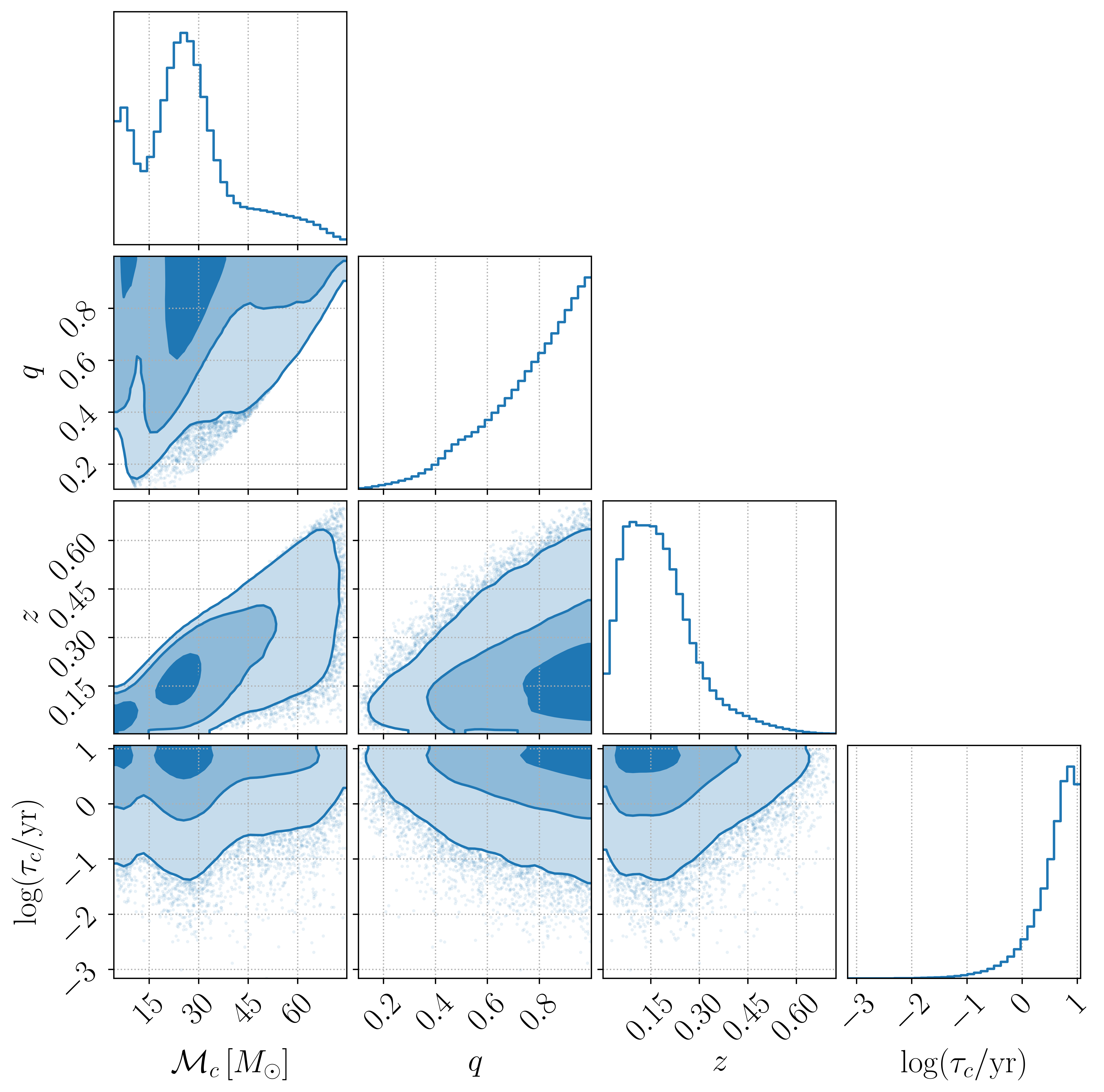}
    \caption{Probability density distribution of the chirp mass, mass ratio, redshift, and logarithm of the time to coalescence for SBBHs detected by LGWA for 10-year observation.}
    \label{fig:LGWA_param_dist}
\end{figure}

We next examine the parameter distributions of the events detectable by LGWA. Fig.~\ref{fig:LGWA_param_dist} shows the joint probability density contours for the chirp mass, mass ratio, redshift, and $\log(\tau_c/{\rm yr})$. In addition to the dominant chirp-mass peak near ${\cal M}_c\simeq30\,M_\odot$, a secondary peak appears at ${\cal M}_c\simeq10\,M_\odot$, corresponding to the lower-mass Gaussian component of the adopted mass model. As in the LISA-like detector case, the mass-ratio distribution favors equal-mass binaries with $q\simeq1$. LGWA’s enhanced sensitivity to low-mass systems makes the secondary chirp-mass peak more prominent.
But unlike the distributions obtained for LISA-like detectors, the redshift distribution for LGWA peaks near $z\simeq0.1$ and decreases rapidly toward $z=0$, reflecting the increase in the available source volume with redshift. Finally, the $\tau_c$ distribution peaks near the assumed observation duration and exhibits a sharp cutoff at $T_{\rm obs}=10\,{\rm yr}$. This behavior is expected because, as discussed in Sec.~\ref{sec:detectability}, the SNR decreases sharply once $\tau_c$ exceeds $T_{\rm obs}$.

\begin{figure*}
    \centering
    \includegraphics[width=\linewidth]{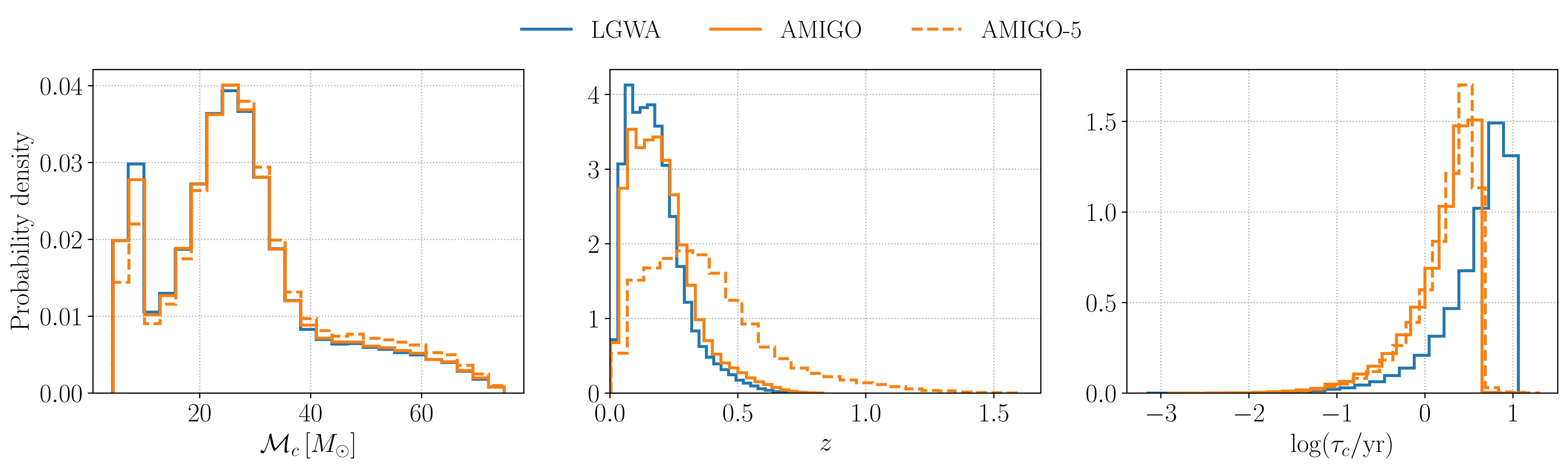}
    \caption{Probability density distribution of the chirp mass, redshift, and logarithm of the time to coalescence for SBBHs detected by decihertz GW observatories LGWA for 10-year observation and AMIGO/AMIGO-5 for 4-year observation, respectively.}
    \label{fig:deci_param_dist}
\end{figure*}

We further compare the one-dimensional parameter distributions of the events detectable by LGWA and AMIGO, as shown in Fig.~\ref{fig:deci_param_dist}. Their chirp mass and redshift distributions are very similar because the two detectors have comparable detection reaches, as illustrated in Fig.~\ref{fig:SNR_contour}. The main difference lies in the $\tau_c$ distribution: for AMIGO, it peaks near $\tau_c\simeq4\,{\rm yr}$, corresponding to its assumed observation duration. Thus, the $\tau_c$ distributions for LGWA and AMIGO follow the same general pattern, with each peaking near its respective value of $T_{\rm obs}$. In addition, we examine how the AMIGO-5 design affects the detected parameter distributions. We find that the chirp-mass and $\tau_c$ distributions remain similar to those obtained for the original AMIGO design, whereas the redshift distribution extends to $z\simeq1.6$ owing to the increased detection reach.

\section{Multiband detection of SBBHs}
\label{sec:multi_band}

Having examined the detection prospects for individual frequency bands, we now turn to forecasts of multiband detection yields in this section.

For joint observations with millihertz and decihertz detectors, detectability depends on the relative timing of their observing periods. Multiband detections are possible when the two periods overlap or when the decihertz observation begins sufficiently soon after the millihertz observation ends. We denote the time delay between the two observing periods by $T_{\rm delay}$ and define $T_{\rm delay}=0$ when the decihertz observation begins immediately after the millihertz observation ends. Under this convention, $T_{\rm delay}<0$ indicates an overlap between the observing periods, whereas $T_{\rm delay}>0$ indicates a gap between them. The minimum time to coalescence required for multiband detection is therefore 
\begin{equation}
    \tau_{c,\min}^{\rm multi} =\max(T_{\rm delay},0), 
    \label{eq:tau_c_min}
\end{equation}
which reduces to zero when the observing periods overlap.

On the other hand, as discussed in Sec.~\ref{sec:detectability}, most of the SNR in the decihertz band accumulates during the final few days before merger—a timescale much shorter than $T_{\rm obs}$. Therefore, a source detected in the millihertz band must evolve sufficiently close to coalescence during the decihertz observing period for its SNR to exceed the decihertz detection threshold. Such a source will merge shortly thereafter and can subsequently be observed by ground-based detectors. The approximate upper bound on the time to coalescence is thus given by 
\begin{equation}
    \tau_{c,\max}^{\rm multi}=T_{\rm obs}^{\rm milli}+T_{\rm delay}+T_{\rm obs}^{\rm deci}.
    \label{eq:tau_c_max}
\end{equation}
Sources with $\tau_c>\tau_{c,\max}^{\rm multi}$ may still be detectable in the millihertz band, but they will not evolve sufficiently close to coalescence before the decihertz observing period ends and will therefore remain below the decihertz detection threshold. They could nevertheless be observed later by ground-based detectors if the waiting time is sufficiently long, as discussed in Ref.~\cite{Gerosa:2019dbe}. In summary, the effective time-to-coalescence window for three-band detections is approximately
\begin{equation}
    \max(T_{\rm delay},0)<\tau_c^{\rm multi}<T_{\rm obs}^{\rm millihertz}
    +T_{\rm delay}+T_{\rm obs}^{\rm decihertz}.
\end{equation}

We now move on to forecasting the number of multiband detections. We continue to use Eq.~\eqref{eq:N_space} and the population distribution models introduced in Sec.~\ref{sec:population}, but impose the more restrictive time window described in the previous paragraph. Denoting the two crossover points in $\tau_c$ by $\tau_{\rm th,1}<\tau_{\rm th,2}$, the expected number of multiband detections becomes
\begin{align}
N ={}& \iiint \mathrm{d}m_1\,\mathrm{d}m_2\,\mathrm{d}z\,
p(m_1,m_2)\,\mathcal{R}(z)\,
\frac{\mathrm{d}V_c(z)}{\mathrm{d}z}\,
\frac{1}{1+z}
\nonumber\\
&\times
\max\!\Bigl\{
0,\,
\min\!\left[
\tau_{\rm th,2}(m_1,m_2,z),
\tau_{c,\max}^{\rm multi}
\right]
\nonumber\\
&\hspace{4.2em}
-
\max\!\left[
\tau_{\rm th,1}(m_1,m_2,z),
\tau_{c,\min}^{\rm multi}
\right]
\Bigr\}.
\label{eq:N_space_multi}
\end{align}
with $\tau_{c,\min}^{\rm multi}$ and $\tau_{c,\max}^{\rm multi}$ defined in Eq.~\eqref{eq:tau_c_min} and \eqref{eq:tau_c_max}. For a given value of $T_{\rm delay}$, we first perform a Monte Carlo integration to calculate the expected number of detections by the millihertz detector with Eq.~\eqref{eq:N_space_multi}. For samples that exceed the SNR threshold in the millihertz band, we then calculate their SNRs in the decihertz band and retain only those that also exceed the corresponding threshold. Since ground-based detectors have a substantially greater detection range as shown in Fig.~\ref{fig:SNR_contour}, SBBH events detectable in both the millihertz and decihertz bands are therefore expected to be detectable by ground-based detectors as well on a sky-averaged basis. This procedure yields the expected number of SBBH events observable across all three frequency bands.

\begin{figure*}[t]
    \centering
    \begin{subfigure}[b]{0.49\textwidth}
        \centering
        \includegraphics[width=\linewidth]{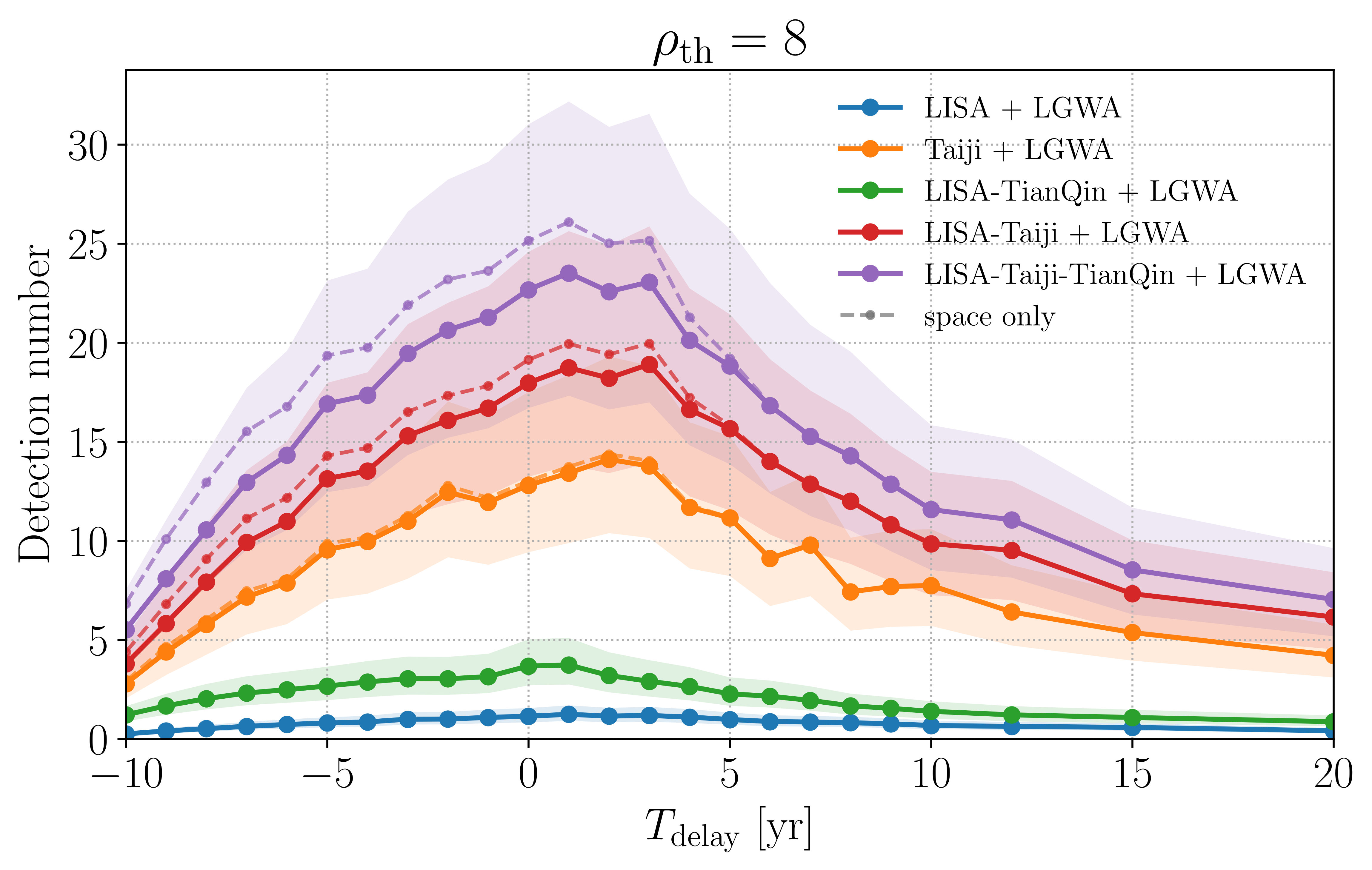}
    \end{subfigure}
    \hfill
    \begin{subfigure}[b]{0.49\textwidth}
        \centering
        \includegraphics[width=\linewidth]{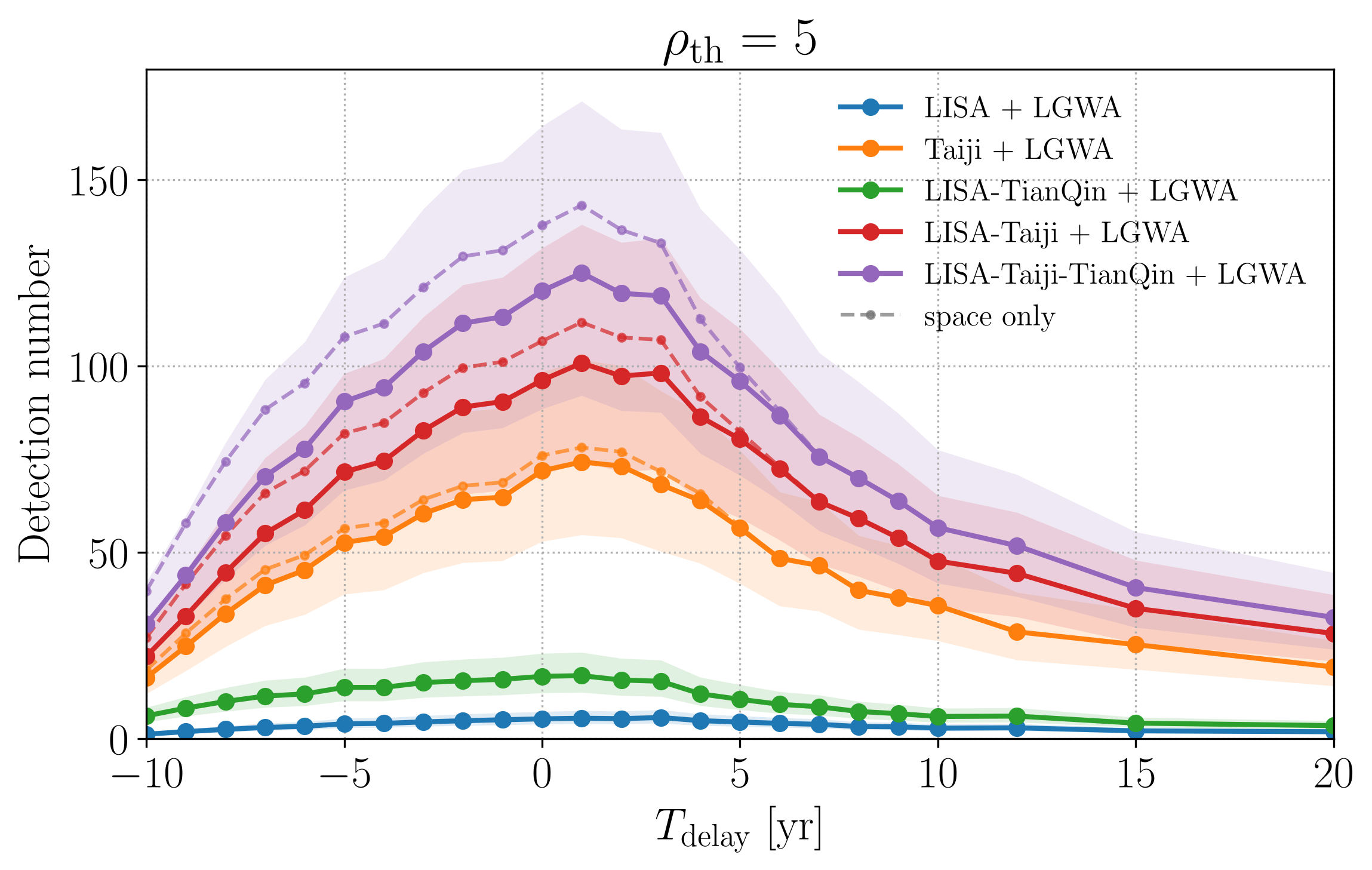}
    \end{subfigure}
    \caption{Expected number of multiband detections by different networks of space-based millihertz detectors and LGWA as a function of $T_{\rm delay}$. The left panel adopts an SNR threshold of $\rho_{\rm th}=8$ in both the millihertz and decihertz bands, while the right panel adopts $\rho_{\rm th}=5$, corresponding to subthreshold searches informed by ground-based detections.}
    \label{fig:detection_Tdelay_lgwa}
\end{figure*}

A further distinction between multiband and single-band detection is that the expected number of multiband detections depends on $T_{\rm delay}$. To identify the scenarios that maximize the detection yield, we therefore perform forecasts over a range of $T_{\rm delay}$ values. Fig.~\ref{fig:detection_Tdelay_lgwa} shows the mean and the $90\%$ confidence interval for the number of multiband SBBH detections by different combinations of millihertz space-based detectors and LGWA. These combinations also include either the LVK or CE ground-based detector network; however, the choice of ground-based detector generation does not affect the predicted detection yield.
For an SNR threshold of $\rho_{\rm th}=8$, shown in the left panel of Fig.~\ref{fig:detection_Tdelay_lgwa}, Taiji and networks that include it can detect several times more events than LISA or TianQin, consistent with the single-band results. More importantly, the number of detections peaks at approximately $T_{\rm delay}=0\text{--}3\,{\rm yr}$, and decreases as $T_{\rm delay}$ becomes either negative or greater than $3\,{\rm yr}$. This trend is independent of the specific detectors constituting the multiband network.

Moreover, Fig.~\ref{fig:detection_Tdelay_lgwa} shows that, for Taiji, the LISA–Taiji network, and the LISA–Taiji–TianQin network, the expected number of detections by the space-based detectors within $\tau_c^{\rm multi}$ exceeds the number of multiband detections that also satisfy the LGWA detection criterion. This result can be understood from Fig.~\ref{fig:SNR_contour}, which shows that Taiji has a slightly greater detection range than LGWA for systems with $m_1\gtrsim 80M_\odot$. Combining Taiji with LISA and TianQin further extends the detection range, resulting in additional events that are detectable by the space-based network but not by LGWA.

Furthermore, it has been shown that information from SBBH signals detected by ground-based observatories can be used to recover the corresponding early-inspiral signals that lie below the conventional SNR threshold in the millihertz or decihertz bands \cite{Wong:2018uwb,Iacovelli:2025kwn}. We therefore also consider a reduced SNR threshold of $\rho_{\rm th}=5$ for the space-based detectors and LGWA, assuming that the late-inspiral signals are independently detected by ground-based observatories. As shown in the right panel of Fig.~\ref{fig:detection_Tdelay_lgwa}, the dependence of the detection yield on $T_{\rm delay}$ remains qualitatively unchanged from the $\rho_{\rm th}=8$ case for all detector networks. However, lowering the threshold increases the expected number of multiband detections by a factor of $4\text{--}5$. The fractional difference between the number of space-based detections and the number of multiband detections also increases when $\rho_{\rm th}=5$, indicating that subthreshold signal recovery is more effective in the millihertz band than in the decihertz band.

\begin{figure*}
    \centering
    \begin{subfigure}[b]{0.49\textwidth}
        \centering
        \includegraphics[width=\linewidth]{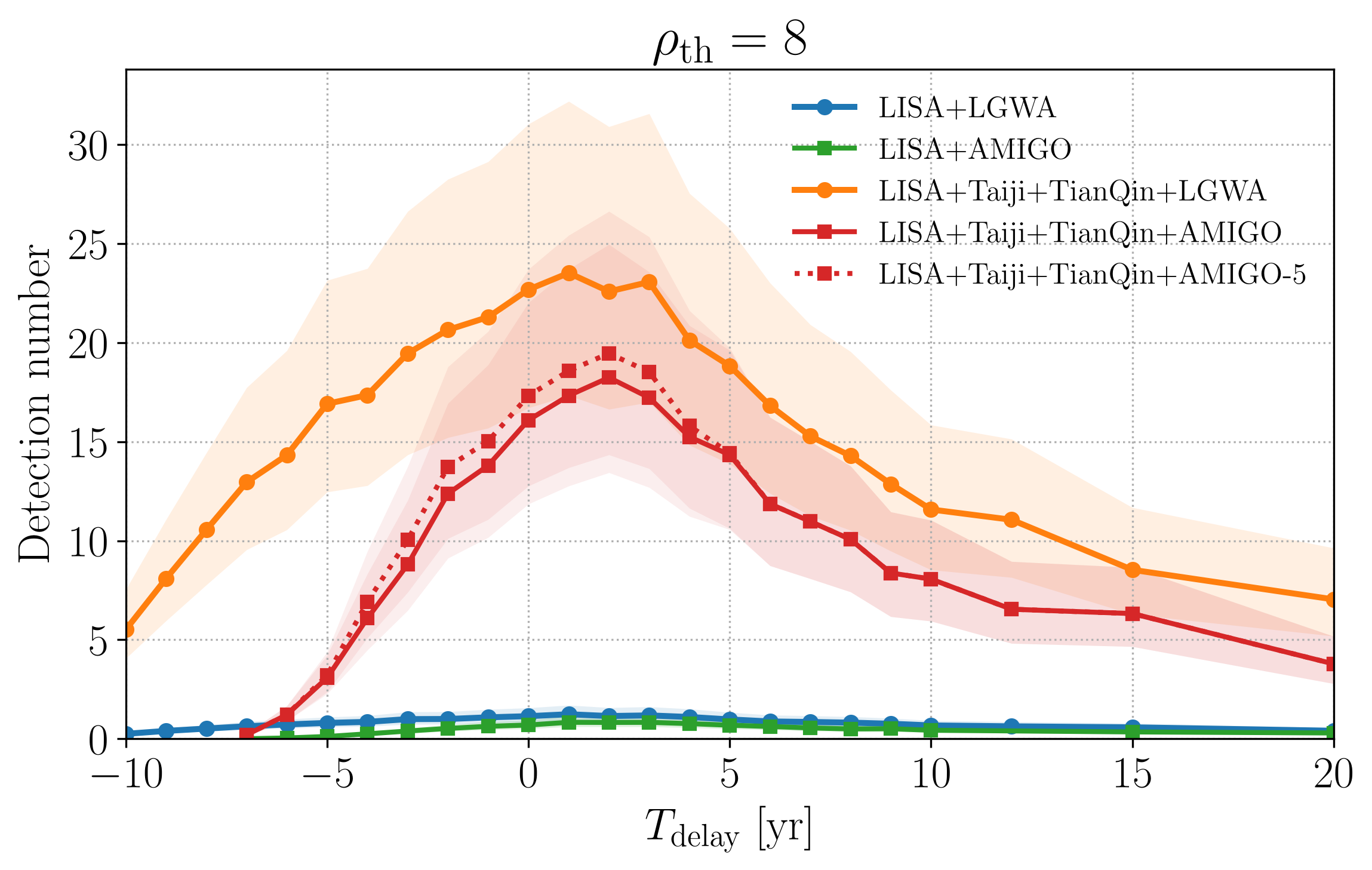}
    \end{subfigure}
    \hfill
    \begin{subfigure}[b]{0.49\textwidth}
        \centering
        \includegraphics[width=\linewidth]{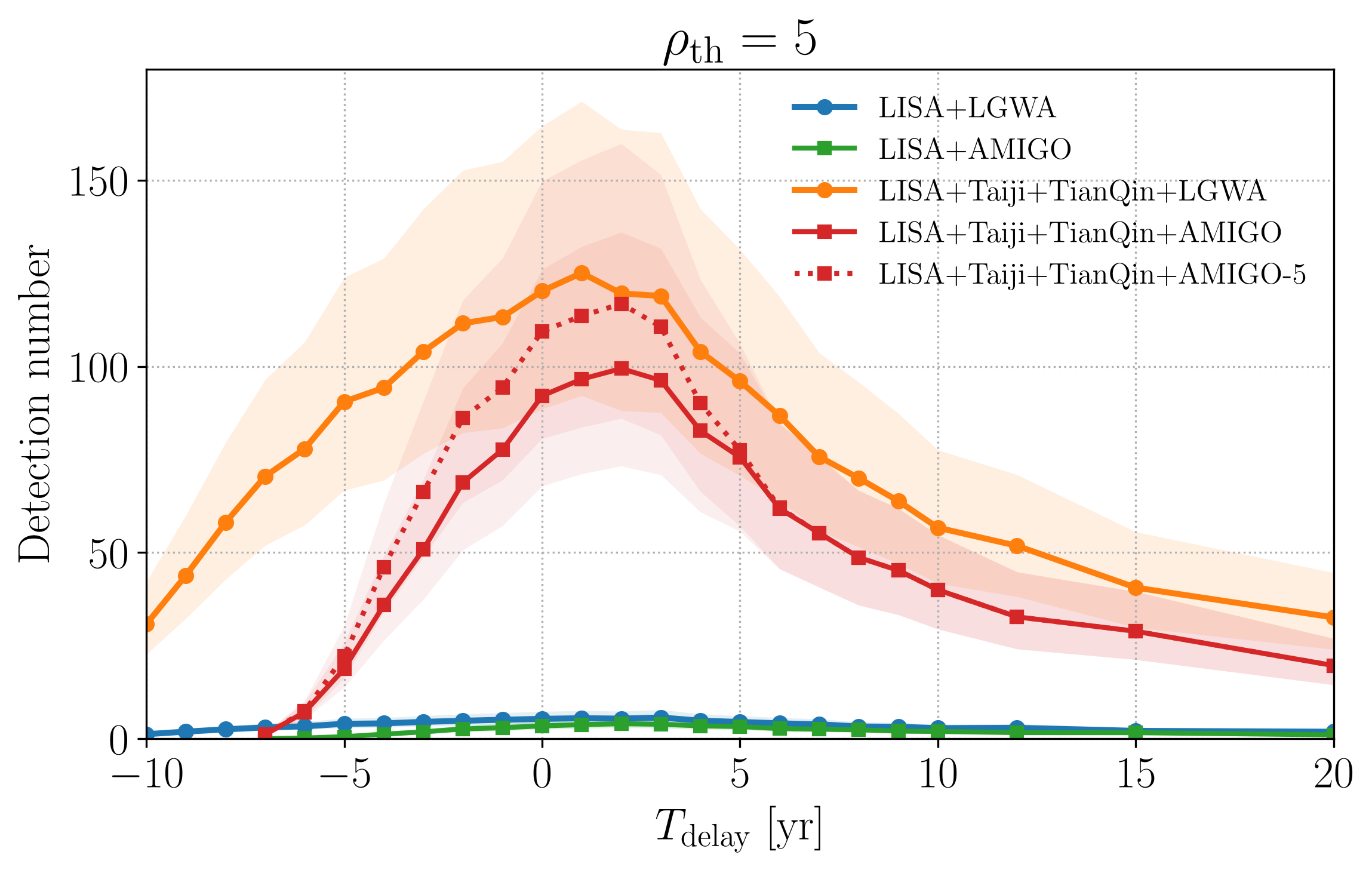}
    \end{subfigure}
    \caption{Expected number of multiband detections by networks combining LISA or LISA--Taiji--TianQin with AMIGO or AMIGO-5, comparing to those with LGWA, as a function of $T_{\rm delay}$. The left and right panels adopt SNR thresholds of $\rho_{\rm th}=8$ and $\rho_{\rm th}=5$, respectively, as described in Fig.~\ref{fig:detection_Tdelay_lgwa}.}
    \label{fig:detection_Tdelay_lgwa_amigo}
\end{figure*}

For comparison, we also investigate the expected number of multiband detections and its dependence on $T_{\rm delay}$ for networks comprising space-based millihertz detectors and the space-based decihertz detector AMIGO/AMIGO-5. Fig.~\ref{fig:detection_Tdelay_lgwa_amigo} compares these results with those obtained using LGWA as the decihertz detector. For clarity, we show only the LISA + decihertz and LISA–Taiji–TianQin + decihertz networks in the figure. Results for the remaining detector combinations are presented in Table~\ref{tab:multiband_number}, which summarizes the maximum expected number of multiband detections over the considered range of $T_{\rm delay}$ for possible detector networks.

Fig.~\ref{fig:detection_Tdelay_lgwa_amigo} shows that the number of multiband detections with AMIGO exhibits a dependence on $T_{\rm delay}$ similar to that obtained with LGWA. The detection yield peaks at approximately $T_{\rm delay}=2\,{\rm yr}$ and decreases as $T_{\rm delay}$ shifts in either direction. The LISA + AMIGO network produces a detection yield comparable to that of LISA + LGWA, whereas the LISA–Taiji–TianQin + AMIGO network detects substantially fewer events than its LGWA counterpart.

Because AMIGO has an observation duration of $T_{\rm obs}=4\,{\rm yr}$, its multiband detection yield approaches zero for $T_{\rm delay}\lesssim-7\,{\rm yr}$. In this regime, AMIGO begins operating more than approximately $3\,{\rm yr}$ before the millihertz space-based detectors, leaving less than one year of overlapping observations. Consequently, only systems with $\tau_c<1\,{\rm yr}$ can be observed in both frequency bands. In this respect, LGWA provides a longer multiband observation window than AMIGO.

Finally, Fig.~\ref{fig:detection_Tdelay_lgwa_amigo} shows that networks combining Taiji with AMIGO-5 detect more multiband events than the corresponding networks containing the original AMIGO, owing to the greater detection range of AMIGO-5. By contrast, we find that for networks containing only LISA or LISA and TianQin in the millihertz band, the difference between AMIGO-5 and AMIGO is negligible. On the other hand, when the SNR threshold is lowered to $\rho_{\rm th}=5$, the improvement provided by AMIGO-5 becomes substantially greater than that for $\rho_{\rm th}=8$. As shown in the right panel of Fig.~\ref{fig:detection_Tdelay_lgwa_amigo}, the maximum multiband detection yield of the network LISA--Taiji--TianQin + AMIGO-5 becomes comparable to that of the corresponding LGWA networks. Therefore, with subthreshold signal searches informed by ground-based detections and an optimal multiband observation schedule, AMIGO-5 can achieve a multiband detection yield comparable to that of LGWA, despite its shorter observation period.
\begin{table*}[]
    \centering
    \begin{tabular}{c|cccccc}
        \hline
        \multirow{2}{*}{\makecell{Decihertz\\detector}} & \multicolumn{6}{c}{Millihertz space-based detector(s)} \\
        \cline{2-7}
         & LISA & Taiji & TianQin & LISA--Taiji & LISA--TianQin & LISA--Taiji--TianQin \\
        \hline
        LGWA & $1.2^{+0.5}_{-0.3}\,(5.7^{+2.1}_{-1.5})$ & $~14.1^{+5.2}_{-3.7}\,(74.3^{+27.4}_{-19.6})$ & $~1.2^{+0.5}_{-0.3}\,(5.4^{+2.0}_{-1.4})$ & $~18.9^{+7.0}_{-5.0}\,(100.9^{+37.1}_{-26.6})$ & $~3.7^{+1.4}_{-1.0}\,(17.0^{+6.3}_{-4.5})$ & $23.5^{+8.7}_{-6.2}\,(125.1^{+46.1}_{-32.9})$ \\
        AMIGO & $0.8^{+0.3}_{-0.2}\,(4.1^{+1.5}_{-1.1})$ & $~9.9^{+3.6}_{-2.6}\,(55.8^{+20.6}_{-14.7})$ & $~1.0^{+0.4}_{-0.3}\,(4.3^{+1.6}_{-1.1})$ & $~14.0^{+5.2}_{-3.7}\,(79.1^{+29.1}_{-20.8})$ & $~2.6^{+1.0}_{-0.7}\,(13.2^{+4.9}_{-3.5})$ & $~18.2^{+6.7}_{-4.8}\,(99.4^{+36.6}_{-26.2})$ \\
        AMIGO-5 & $0.8^{+0.3}_{-0.2}\,(4.1^{+1.5}_{-1.1})$ & $~9.9^{+3.6}_{-2.6}\,(56.4^{+20.8}_{-14.8})$ & $~1.0^{+0.4}_{-0.3}\,(4.3^{+1.6}_{-1.1})$ & $~14.4^{+5.3}_{-3.8}\,(88.0^{+32.4}_{-23.2})$ & $~2.6^{+1.0}_{-0.7}\,(13.2^{+4.9}_{-3.5})$ & $~19.5^{+7.2}_{-5.1}\,(116.8^{+43.0}_{-30.7})$ \\
        \hline
    \end{tabular}
    \caption{Maximum expected multiband detections over the considered range of $T_{\rm delay}$ for different configurations of millihertz and decihertz detectors combined with either LVK or 2-CE ground-based detectors, assuming SNR thresholds of $\rho_{\rm th}=8$ ($\rho_{\rm th}=5$ for ground-informed searches).}
    \label{tab:multiband_number}
\end{table*}

\section{Parameter estimation}
\label{sec:pe}

Having investigated the expected number of multiband detections in the previous section, we now turn to parameter estimation for multiband events and assess the improvement achieved relative to single-band observations in this section. We illustrate the parameter-estimation uncertainties using the Fisher-matrix approximation for a nearby GW150914-like event. The injected parameters are $m_1=36.0\,M_\odot$, $m_2=29.0\,M_\odot$, $D_L=200\,{\rm Mpc}$, inclination angle $\iota=0.9$, $\tau_c=4\,{\rm yr}$, coalescence phase $\phi_c=1.5$, component spin $a_1=0.1$, $a_2=0.05$, right ascension and declination $(3.0, 0.0)$, and polarization angle $\psi=0.4$.

The Fisher-matrix approximation is widely used to forecast uncertainties in gravitational-wave parameter estimation \cite{Finn:1992wt,Cutler:1994ys,Poisson:1995ef}. Assuming stationary Gaussian noise and sufficiently high event SNR, the likelihood can be approximated as a multivariate Gaussian near its maximum. For a waveform $h(\boldsymbol{\theta})$ characterized by parameters $\boldsymbol{\theta}$, the Fisher information matrix is defined as $\Gamma_{ij}=(\partial_i h|\partial_j h)$, where $(\cdot|\cdot)$ denotes the noise-weighted inner product defined by
\begin{equation}
    (h|g) = \int_{f_{\rm min}}^{f_{\rm max}}
\frac{\tilde{h}^{*}(f)\tilde{g}(f)+\tilde{h}(f)\tilde{g}^{*}(f)}{S_n(f)}\,{\rm d}f.
\end{equation}
The corresponding covariance matrix is approximated by the inverse Fisher matrix, $\Sigma=\Gamma^{-1}$, and the marginalized $1\sigma$ uncertainty in parameter $\theta_i$ is given by $\sigma_i=\sqrt{\Sigma_{ii}}$. For multiband observations, the information from statistically independent detectors and frequency bands can be combined by summing their individual Fisher matrices, allowing us to quantify the improvement in parameter estimation relative to single-band measurements.

\subsection{Multiband network with LGWA}

\begin{figure*}
    \centering
    \begin{subfigure}[b]{0.54\textwidth}
        \centering
        \includegraphics[width=\linewidth]{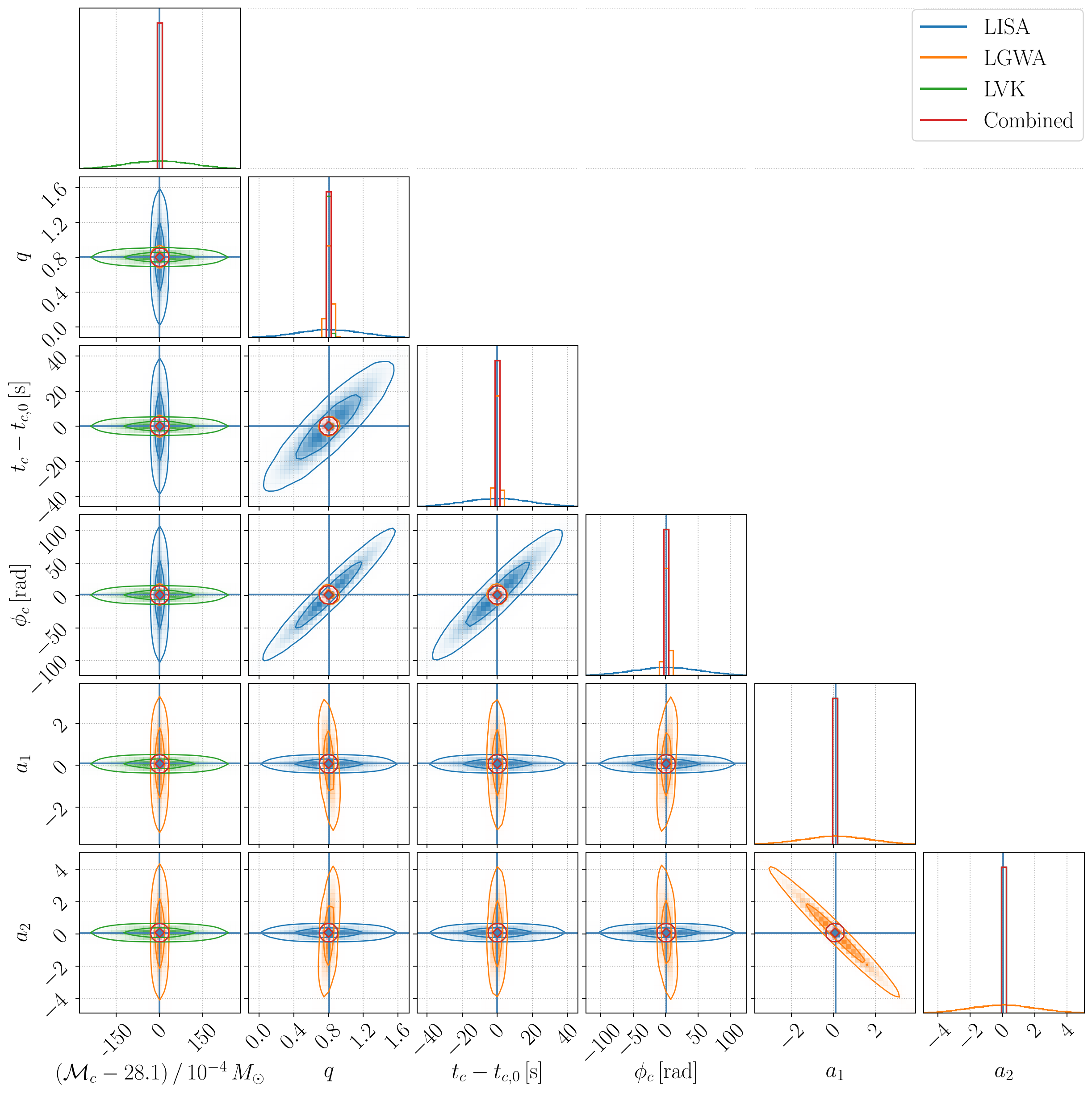}
    \end{subfigure}
    \hfill
    \begin{subfigure}[b]{0.45\textwidth}
        \centering
        \includegraphics[width=\linewidth]{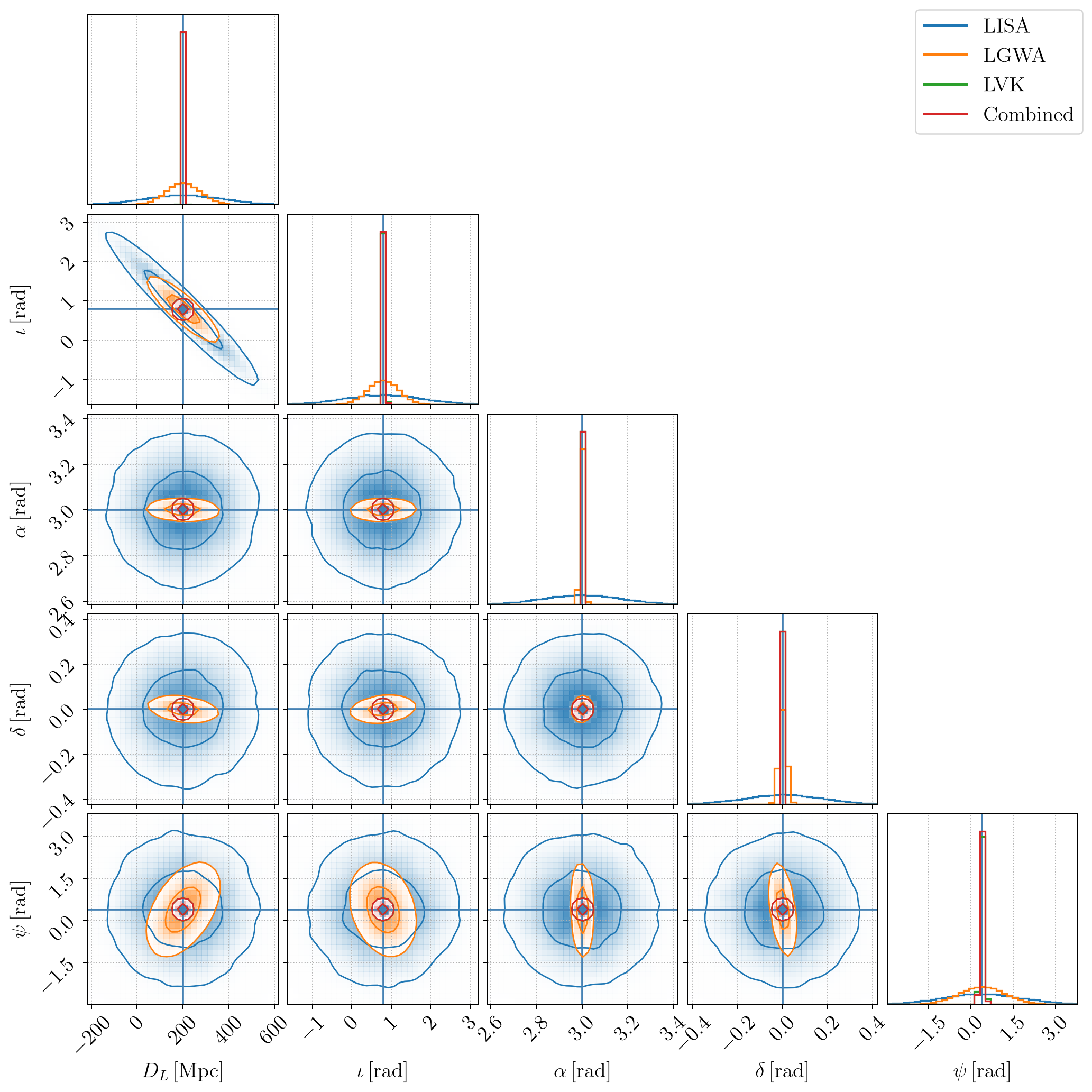}
    \end{subfigure}
    \caption{Corner plots showing the constraints on all waveform parameters of the GW150914-like event obtained from Fisher-matrix analysis using LISA, LGWA, and LVK individually, as well as their combined three-band observations. The left panel shows the intrinsic parameters, while the right panel shows the extrinsic parameters.}
    \label{fig:gwfish_corner}
\end{figure*}

We compute the Fisher matrix with respect to all injected waveform parameters using the \texttt{GWFish} package and the IMRPhenomXPHM waveform model \cite{Pratten:2020ceb}, which incorporates higher-order modes that help break the $D_L$–$\iota$ degeneracy. Fig.~\ref{fig:gwfish_corner} presents corner plots of the resulting parameter constraints for the GW150914-like event, obtained using LISA, LGWA, and LVK individually, as well as their combined three-band observations. The LVK network considered here comprises the LIGO Livingston and Hanford detectors, Virgo, and KAGRA, all operating at their design sensitivities. The SNRs of the event in LISA, LGWA, and LVK are $5.6$, $84.8$, and $313.8$, respectively. Consequently, most parameters are constrained most tightly by LVK, followed by LGWA and then LISA, with one notable exception: the chirp mass is best constrained by LGWA, followed by LISA and LVK. This is because the signal remains in the inspiral regime for an extended period in the LGWA and LISA bands, allowing the accumulated phase evolution to place stringent constraints on the chirp mass. Moreover, LGWA observes the final few days of the inspiral, during which the phase evolves more rapidly, and the signal frequency exceeds the upper limit of the LISA band. LGWA therefore provides a tighter constraint on the chirp mass than LISA.

We also summarize the $1\sigma$ uncertainties in all waveform parameters for individual detectors and multiband networks comprising LGWA, space-based millihertz detectors, and ground-based detectors in Table~\ref{tab:LGWA_uncertainty}. Relative uncertainties are reported for the detector-frame chirp mass $\mathcal{M}_c$ and luminosity distance $D_L$. In addition to LVK, we investigate parameter constraints from a next-generation ground-based detector network, represented in this work by two CEs located at the LIGO Livingston and Hanford sites. Although future networks may comprise configurations such as ET together with one or two CEs, we do not examine each configuration separately. Instead, we adopt the 2-CE network as a representative example. Moreover, assuming a bivariate Gaussian distribution for the sky-position posterior, we calculate the $90\%$ confidence sky-localization area as
\begin{equation}
    \Delta\Omega = -2\pi\ln(1-0.9)\,|\cos\theta_{\rm dec}|\,
\sqrt{\sigma_{\rm RA}^{2}\sigma_{\rm dec}^{2} -\Sigma_{\rm RA,dec}^{2}},
\end{equation}
with the uncertainty of right ascension and declination, and the covariance between them. 
\begin{table*}[]
    \centering
    \begin{tabular}{c|cccccccccc}
    \hline
        \multirow{2}{*}{Detector} & \multicolumn{10}{c}{Parameter uncertainty} \\
        \cline{2-11}
         & $\Delta {\cal M}_c/{\cal M}_c$ & $\Delta q$ & $\Delta D_L/D_L$ & $\Delta \iota$ & $\Delta t_c$ [s] & $\Delta \phi_c$ & $\Delta a_1$ & $\Delta a_2$ & $\Delta\Omega~[{\rm deg}^2]$ & $\Delta \psi$ \\
        \hline
        LISA & $~5.5\times10^{-6}~$ & $0.39$ & $87\%$ & $1.0$ & $19.2$ & $51.9$ & - & - & $1476.0$ & $1.4$ \\
        LGWA & $6.3\times10^{-7}$ & $0.03$ & $38\%$ & $0.4$ & $1.1$ & $3.2$ & - & - & $1.88$ & $0.8$ \\
        LVK & $4.3\times10^{-4}$ & $0.01$ & $2.0\%$ & $0.03$ & $3.1\times10^{-5}$ & $0.1$ & $0.03$ & $0.04$ & $0.065$ & $0.06$ \\
        2-CE & $~7.5\times10^{-6}~$ & $0.001$ & $0.3\%$ & $0.003$ & $~1.9\times10^{-5}~$ & $~0.01~$ & $~0.003~$ & $~0.003~$ & $~0.0087~$ & $~0.007~$ \\
        LGWA+LVK & $1.8\times10^{-7}$ & $0.003$ & $1.9\%$ & $0.02$ & $1.4\times10^{-5}$ & $0.1$ & $0.02$ & $0.03$ & $~0.0011~$ & $0.05$ \\
        LISA+LGWA+LVK & $1.1\times10^{-7}$ & $0.004$ & $1.9\%$ & $0.02$ & $1.4\times10^{-5}$ & $0.1$ & $0.02$ & $0.03$ & $~0.0011~$ & $0.05$ \\
        \makecell{~\\LISA--Taiji--TianQin\\+LGWA+LVK} & $7.8\times10^{-8}$ & $0.002$ & $1.9\%$ & $0.02$ & $1.4\times10^{-5}$ & $0.1$ & $0.02$ & $0.03$ & $~0.0010~$ & $0.05$ \\
        LGWA+2-CE & $7.5\times10^{-8}$ & $0.0007$ & $0.2\%$ & $0.002$ & $1.7\times10^{-6}$ & $0.01$ & $0.002$ & $0.003$ & $~8.3\times10^{-5}~$ & $0.005$ \\
        LISA+LGWA+2-CE & $3.7\times10^{-8}$ & $0.0006$ & $0.2\%$ & $0.002$ & $1.7\times10^{-6}$ & $0.01$ & $0.002$ & $0.003$ & $~8.1\times10^{-5}~$ & $0.005$ \\
        \makecell{~\\LISA--Taiji--TianQin\\+LGWA+2-CE} & $2.8\times10^{-8}$ & $0.0006$ & $0.2\%$ & $0.002$ & $1.7\times10^{-6}$ & $0.01$ & $0.002$ & $0.003$ & $~8.0\times10^{-5}~$ & $0.005$ \\
        \hline
    \end{tabular}
    \caption{Parameter uncertainty forecasted with Fisher matrix for individual detector or detector network consisting of ground-based detectors, LISA-like detectors and LGWA.}
    \label{tab:LGWA_uncertainty}
\end{table*}

First, Table~\ref{tab:LGWA_uncertainty} shows that LISA and LGWA can individually constrain $\mathcal{M}_c$ two to three orders of magnitude more tightly than LVK. However, their constraints on the other parameters are considerably weaker, and they provide essentially no information about the component spins. This is expected because these detectors do not observe the late-inspiral regime, where higher-order post-Newtonian spin effects become increasingly prominent. On the other hand, the 2-CE network constrains $\mathcal{M}_c$ almost as tightly as LISA, although its uncertainty remains approximately one order of magnitude larger than that of LGWA. It also improves the constraints on most other parameters by roughly one order of magnitude relative to LVK.

Second, Table~\ref{tab:LGWA_uncertainty} shows that combining LGWA with LVK significantly reduces the uncertainties in most parameters relative to LVK alone. In particular, the uncertainty in $\mathcal{M}_c$ decreases by more than three orders of magnitude, while the sky-localization area $\Delta\Omega$ improves by more than one order of magnitude. The uncertainties in most other parameters decrease by approximately $20\%\text{--}70\%$, whereas those in $D_L$ and $\phi_c$ remain nearly unchanged. Adding LISA to form the three-band LISA+LGWA+LVK network further reduces the uncertainty in $\mathcal{M}_c$ by approximately $38\%$ relative to LGWA+LVK, but provides negligible improvements for the other parameters. Replacing LISA with the more sensitive LISA--Taiji--TianQin network decreases the uncertainty in $\mathcal{M}_c$ by approximately $57\%$ relative to LGWA+LVK, while still yielding only modest improvements in the remaining parameters. These results demonstrate that decihertz observations play the dominant role in improving most waveform-parameter constraints in multiband analyses. Millihertz observations primarily enhance the measurements of $\mathcal{M}_c$ and $\Delta\Omega$, with limited benefits for the other parameters even when multiple millihertz detectors are combined.

Finally, replacing LVK with the 2-CE network, Table~\ref{tab:LGWA_uncertainty} shows that the addition of LGWA reduces both $\Delta\mathcal{M}_c$ and $\Delta\Omega$ by approximately two orders of magnitude relative to 2-CE observations alone, while producing modest improvements in most other parameter uncertainties. Further including LISA or the LISA--Taiji--TianQin network reduces $\Delta\mathcal{M}_c$ by approximately $51\%$ or $63\%$, respectively, relative to LGWA+2-CE, but yields little additional improvement in the other parameters, consistent with the LVK results. A notable exception is the coalescence time $t_c$. Although LVK and 2-CE observations alone both constrain $t_c$ to uncertainties of order $10^{-5}\,{\rm s}$, their improvements from multiband observations differ substantially. Adding LGWA reduces $\Delta t_c$ by approximately $55\%$ relative to LVK alone, whereas adding it to the 2-CE network reduces $\Delta t_c$ by approximately one order of magnitude. This result indicates that multiband observations provide a particularly significant improvement in the measurement of the coalescence time when combined with next-generation ground-based detectors.

\subsection{Multiband network with AMIGO}

For comparison, we also perform a Fisher-matrix analysis of multiband networks in which LGWA is replaced by the decihertz detector AMIGO. The resulting parameter uncertainties are presented in Table~\ref{tab:AMIGO_uncertainty}. In general, AMIGO alone constrains most parameters slightly more tightly than LGWA, with two notable exceptions. First, AMIGO yields a slightly larger uncertainty in $\mathcal{M}_c$. This is because LGWA has a lower low-frequency cutoff and can therefore observe the inspiral signal for a longer duration, accumulating more phase information. Second, AMIGO constrains $t_c$ substantially more tightly than LGWA. Its higher upper-frequency cutoff allows it to observe the binary closer to coalescence, thereby providing a more precise measurement of the coalescence time. However, AMIGO alone is still unable to place meaningful constraints on the component spins. For multiband networks combining AMIGO with ground-based detectors, including three-band configurations, replacing LGWA with AMIGO produces only minor differences in the parameter uncertainties, because ground-based detectors provide most of the constraining power for the majority of waveform parameters.
\begin{table*}[]
    \centering
    \begin{tabular}{c|cccccccccc}
    \hline
        \multirow{2}{*}{Detector} & \multicolumn{10}{c}{Parameter uncertainty} \\
        \cline{2-11}
         & $\Delta {\cal M}_c/{\cal M}_c$ & $\Delta q$ & $\Delta D_L/D_L$ & $\Delta\iota$ & $\Delta t_c$ [s] & $\Delta\phi_c$ & $\Delta a_1$ & $\Delta a_2$ & $\Delta\Omega~[{\rm deg}^2]$ & $\Delta\psi$ \\
        \hline
        AMIGO & $~7.8\times10^{-7}$ & $~0.035$ & $19\%$ & $0.25$ & $~0.037$ & $2.8$ & - & - & $1.06$ & $0.27$ \\
        AMIGO+LVK & $~2.1\times10^{-7}$ & $~0.005$ & $1.9\%$ & $0.02$ & $~2.3\times10^{-5}$ & $0.10$ & $0.02$ & $0.03$ & $0.05$ & $0.06$ \\
        LISA+AMIGO+LVK & $~1.3\times10^{-7}$ & $~0.003$ & $1.9\%$ & $0.02$ & $~2.3\times10^{-5}$ & $0.10$ & $0.02$ & $0.03$ & $0.05$ & $0.06$ \\
        \makecell{~\\LISA--Taiji--TianQin\\+AMIGO+LVK} & $~8.1\times10^{-8}$ & $~0.003$ & $1.9\%$ & $0.02$ & $~2.3\times10^{-5}$ & $0.10$ & $0.02$ & $0.03$ & $0.05$ & $0.05$ \\
        AMIGO+2-CE & $~6.9\times10^{-8}$ & $~0.0008$ & $0.3\%$ & $0.003$ & $~1.8\times10^{-5}$ & $~0.01$ & $~0.002$ & $~0.003$ & $0.008$ & $0.006$ \\
        LISA+AMIGO+2-CE & $~4.3\times10^{-8}$ & $~0.0007$ & $0.3\%$ & $0.003$ & $~1.8\times10^{-5}$ & $~0.01$ & $~0.002$ & $~0.003$ & $0.008$ & $0.006$ \\
        \makecell{~\\LISA--Taiji--TianQin\\+AMIGO+2-CE} & $~3.3\times10^{-8}$ & $~0.0007$ & $0.3\%$ & $0.003$ & $~1.8\times10^{-5}$ & $~0.01$ & $~0.002$ & $~0.003$ & $0.008$ & $0.006$ \\
        \hline
    \end{tabular}
    \caption{Parameter uncertainty forecasted with Fisher matrix for individual detector or detector network consisting of ground-based detectors, LISA-like detectors and AMIGO.}
    \label{tab:AMIGO_uncertainty}
\end{table*}

Furthermore, we repeat the analysis after replacing AMIGO with the more sensitive AMIGO-5 configuration in the multiband networks. The resulting parameter uncertainties are presented in Table~\ref{tab:AMIGO-5_uncertainty}. When operating alone, AMIGO-5 reduces the uncertainties in most waveform parameters by approximately a factor of two relative to AMIGO, while providing little improvement in the sky-localization area. In multiband observations, however, the improvements are generally modest because AMIGO-5 is less sensitive than AMIGO at high frequencies, where much of the information about most waveform parameters is obtained. The most substantial improvement is found for the chirp mass: depending on the multiband network, $\Delta\mathcal{M}_c$ decreases by approximately $24\%\text{--}61\%$ relative to the corresponding network containing AMIGO. This improvement arises from the greater low-frequency sensitivity of AMIGO-5, which allows it to accumulate more information during the long inspiral phase. Therefore, the primary parameter-estimation benefit of increasing the AMIGO arm length is an improved measurement of the source masses, particularly the chirp mass.
\begin{table*}[]
    \centering
    \begin{tabular}{c|cccccccccc}
    \hline
        \multirow{2}{*}{Detector} & \multicolumn{10}{c}{Parameter uncertainty} \\
        \cline{2-11}
         & $\Delta {\cal M}_c/{\cal M}_c$ & $\Delta q$ & $\Delta D_L/D_L$ & $\Delta\iota$ & $\Delta t_c$ [s] & $\Delta\phi_c$ & $\Delta a_1$ & $\Delta a_2$ & $\Delta\Omega~[{\rm deg}^2]$ & $\Delta \psi$ \\
        \hline
        AMIGO-5 & $~1.8\times10^{-7}$ & $~0.007$ & $8.5\%$ & $0.11$ & $0.028$ & $0.47$ & - & - & $1.01$ & $0.15$ \\
        AMIGO-5+LVK & $~8.2\times10^{-8}$ & $~0.002$ & $1.9\%$ & $0.02$ & $~2.5\times10^{-5}$ & $0.10$ & $0.02$ & $0.03$ & $0.06$ & $0.05$ \\
        LISA+AMIGO-5+LVK & $~7.3\times10^{-8}$ & $~0.002$ & $1.9\%$ & $0.02$ & $~2.5\times10^{-5}$ & $0.10$ & $0.02$ & $0.03$ & $0.06$ & $0.05$ \\
        \makecell{~\\LISA--Taiji--TianQin\\+AMIGO-5+LVK} & $~5.8\times10^{-8}$ & $~0.002$ & $1.9\%$ & $0.02$ & $~2.5\times10^{-5}$ & $0.10$ & $0.02$ & $0.03$ & $0.06$ & $0.05$ \\
        AMIGO-5+2-CE & $~3.4\times10^{-8}$ & $~0.0006$ & $0.3\%$ & $0.003$ & $~1.8\times10^{-5}$ & $~0.01$ & $~0.002$ & $~0.003$ & $0.008$ & $0.006$ \\
        LISA+AMIGO-5+2-CE & $~3.1\times10^{-8}$ & $~0.0006$ & $0.3\%$ & $0.003$ & $~1.8\times10^{-5}$ & $~0.01$ & $~0.002$ & $~0.003$ & $0.008$ & $0.006$ \\
        \makecell{~\\LISA--Taiji--TianQin\\+AMIGO-5+2-CE} & $~2.5\times10^{-8}$ & $~0.0005$ & $0.3\%$ & $0.003$ & $~1.8\times10^{-5}$ & $~0.01$ & $~0.002$ & $~0.003$ & $0.008$ & $0.006$ \\
        \hline
    \end{tabular}
    \caption{Parameter uncertainty forecasted with Fisher matrix for individual detector or detector network consisting of ground-based detectors, LISA-like detectors and AMIGO-5.}
    \label{tab:AMIGO-5_uncertainty}
\end{table*}

\section{Conclusions}
\label{sec:conclusion}

We have investigated the prospects for observing SBBH coalescences across the millihertz, decihertz, and ground-based gravitational-wave bands. Our analysis combines population forecasts based on the latest GWTC-4 binary-black-hole mass and merger-rate models with Fisher-matrix estimates for a representative GW150914-like source. We considered the millihertz observatories LISA, Taiji, and TianQin; the decihertz concepts LGWA, AMIGO, and AMIGO-5; and both current-generation LVK and next-generation Cosmic Explorer ground-based networks.

For single-band millihertz observations with a four-year mission and a SNR threshold of $\rho_{\rm th}=8$, we find expected yields of $4.6^{+1.7}_{-1.2}$, $29.6^{+10.9}_{-7.8}$, and $1.1^{+0.4}_{-0.3}$ SBBHs for LISA, Taiji, and TianQin, respectively. Combining the three observatories increases the expected yield to $53.6^{+19.7}_{-14.1}$. The detected population is preferentially composed of nearly equal-mass systems with detector-frame chirp masses near the high-mass peak of the adopted population model. Decihertz detectors probe a complementary stage of the inspiral and, because their sources are generally also within the reach of ground-based observatories, provide the bridge required for three-band observations.

The number of multiband detections depends not only on sensitivity but also on the relative observing schedules. The substantially shorter range of event merger times compatible with multiband observations results in fewer detections than in the corresponding single-band observations. For all detector combinations studied here, the yield is largest when the decihertz mission begins approximately $0$--$3\,\mathrm{yr}$ after the millihertz observations end. At $\rho_{\rm th}=8$, the maximum detections over $T_{\rm delay}$ is $1.2^{+0.5}_{-0.3}$ for the multiband network of LISA, LGWA, and ground-based detectors, which is consistent with previous forecasts for the LISA + LVK network \cite{Gerosa:2019dbe,Buscicchio:2024asl} because the event yield is primarily limited by detectability in the millihertz band. In addition, the largest predicted yield is $23.5^{+8.7}_{-6.2}$ events for LISA--Taiji--TianQin combined with LGWA and ground-based detectors, compared with $18.2^{+6.7}_{-4.8}$ for the corresponding network for AMIGO and $19.5^{+7.2}_{-5.1}$ for AMIGO-5. The longer observing lifetime of LGWA gives it an important advantage when mission overlap is limited. If a confident ground-based detection is used to guide searches for earlier, otherwise subthreshold signals, lowering the millihertz and decihertz threshold to $\rho_{\rm th}=5$ raises the predicted yields by a factor of approximately $4$--$5$. The maximum yields then reach $125.1^{+46.1}_{-32.9}$, $99.4^{+36.6}_{-26.2}$, and $116.8^{+43.0}_{-30.7}$ events for the multiband networks containing LGWA, AMIGO, and AMIGO-5, respectively. Targeted archival searches are therefore likely to be a central component of multiband observing strategies. It should be noted that these maximal detection yields assume continuous operation and should therefore be regarded as optimistic upper limits, as realistic duty cycles—particularly for ground-based detectors and TianQin—will reduce the number of coincident observations. Ground-based detectors experience commissioning, maintenance, and periods of degraded data quality, while TianQin’s observing schedule includes interruptions imposed by its orbital configuration and Sun-avoidance constraints.

Our parameter-estimation results further demonstrate the complementarity of the three frequency regimes. For a GW150914-like binary, millihertz and decihertz observations measure the detector-frame chirp mass much more precisely than LVK because they track a large number of inspiral cycles, while ground-based detectors dominate the constraints on the mass ratio, spins, distance, inclination, and merger phase. Combining LGWA and LVK improves the chirp-mass uncertainty by more than three orders of magnitude and the sky-localization area by more than one order of magnitude, while improving most other parameters by approximately $20$--$70\%$. Adding LISA, or the LISA--Taiji--TianQin network, provides a further improvement of approximately $38\%$ or $57\%$ in the chirp-mass uncertainty, with comparatively small changes in most other parameters. Similar behavior is found with a 2-CE network: the decihertz data primarily improve the chirp mass, localization, and coalescence time, while the millihertz band supplies additional long-baseline phase information. AMIGO generally performs comparably to LGWA in a multiband network, whereas AMIGO-5 chiefly improves the chirp-mass measurement through its enhanced low-frequency sensitivity.

These forecasts establish a clear hierarchy of roles in stellar-mass multiband observations. Ground-based detectors provide the strongest measurement of the merger and of most source parameters; decihertz detectors supply the principal improvement beyond the ground band; and millihertz observatories extend the phase baseline, improve the chirp mass and localization, and provide years of advance warning. The scientific return is therefore determined by the network as a whole rather than by any single detector. Coordinating mission timelines, maintaining sufficient overlap between bands, and developing searches conditioned on later ground-based detections will be crucial for maximizing the number and quality of multiband events. Such observations will enable exceptionally precise studies of binary evolution and black-hole populations, and will strengthen applications to fundamental physics and gravitational-wave cosmology.

\begin{acknowledgments}
A.C. is supported by the China Postdoctoral Science Foundation under Grant No. 2025M773325, and the National Natural Science Foundation of China (NSFC) under Grant No. E414660101, 12147103, W2611007. J. Z. is supported by the scientific research starting grants from the University of Chinese Academy of Sciences (Grant No.~118900M061), the Fundamental Research Funds for the Central Universities (Grants No.~E2EG6602X2 and No.~E2ET0209X2), and the NSFC under Grant No.~12147103. This research was supported by the National Key R\&D Program of China, No. 2025YFE0217300.
\end{acknowledgments}

The data that support the findings of this article are openly available at \cite{data}.

% \appendix
% \section{Forecast with AMIGO-5}

\bibliography{reference}{}
\bibliographystyle{apsrev4-2}

\end{document}